\documentclass[reprint,amsmath,amssymb,aps,nofootinbib]{revtex4-1}

\usepackage{bbold}
\usepackage{color}
\usepackage{graphicx,graphics}
\usepackage{epstopdf}
\usepackage{amsmath}
\usepackage{amssymb}

\DeclareMathAlphabet\mathbfcal{OMS}{cmsy}{b}{n}

\let\a=\alpha \let\be=\beta  \let\g=\gamma  \let\d=\delta \let\e=\varepsilon
     \let\l=\lambda
\let\m=\mu    \let\n=\nu  \let\o=\omega          
\let\s=\sigma \let\t=\tau    \let\ch=\chi
   
\let\G=\Gamma \let\D=\Delta

\let\io=\infty 

\def\eg{{e.g.}}

\def\bD{\mathbf{D}}

\def\beq{\begin{equation}}
\def\eeq{\end{equation}}

\def\DD{\mathcal{D}} \def\OO{\mathcal{O}}  \def\GG{\mathcal{G}}
\def\SS{\mathcal{S}}  \def\LL{\mathcal{L}}

\def\bch{\boldsymbol{\chi}}

\def\bh{\boldsymbol{h}}    \def\bJ{\boldsymbol{J}} 
  \def\bc{\boldsymbol{c}} \def\bG{\boldsymbol{G}} 
\def\bL{\boldsymbol{L}}

\def\calG{\mathcal{G}}

\renewcommand\b[1]{\boldsymbol{#1}} \renewcommand\t[1]{\tilde{#1}}
\newcommand\bt[1]{\boldsymbol{\tilde{#1}}}

\renewcommand\bar[1]{\overline{#1}}

\def\to{\rightarrow}
\def\la{\left\langle}
\def\ra{\right\rangle}
\def\Tr{{\rm Tr }}

\usepackage{graphicx}
\usepackage{dcolumn}
\usepackage{bm}

\def\Id{\mathbb{1}}

\begin{document}

\title{Resummed mean-field inference for strongly coupled data}

\author{Hugo Jacquin}
\affiliation{Laboratoire de Physique Statistique, \'Ecole Normale Sup\'erieure, \\ UMR CNRS 8550, 24 rue Lhomond, 75005 Paris, France}
\author{A. Ran\c{c}on}
\affiliation{
Universit\'e de Lyon, ENS de Lyon, Universit\'e Claude Bernard, CNRS, 
Laboratoire de Physique, F-69342 Lyon, France}

\begin{abstract}
We present a new resummed mean field approximation for inferring the parameters of an Ising or a Potts model from empirical, noisy, 
one- and two-point 
correlation functions. Based on a resummation of a class of diagrams of the small correlation expansion of the log-likelihood, 
the method outperforms standard mean-field inference methods, even when they are regularized. The inference is stable with respect to sampling
noise, contrarily to previous works based either on the small correlation expansion, on the Bethe free energy, or on the mean-field and Gaussian 
models. Because it is mostly analytic, its 
complexity is still very low, requiring an iterative algorithm to solve for $N$ auxiliary variables, that resorts only to matrix inversions and 
multiplications. We test our algorithm on the Sherrington-Kirkpatrick model
submitted to a random external field and large random couplings, and demonstrate that even without regularization, the inference is stable across
the whole phase diagram. In addition, the calculation leads to a consistent estimation of the entropy of the data, 
and allows us to sample form the inferred distribution to obtain artificial data that are consistent with the empirical distribution.
\end{abstract}


\maketitle


\section{Introduction}

In a context of ever increasing data availability, the task of inferring a probability distribution given measured data has become ubiquitous. 
This task is referred to as statistical inference, applications of which can be found in the study of bird flocks \cite{flocks}, finance 
\cite{BRB15,Moro09}, 
neuroscience \cite{TMMABB13,MBFD09,RTH09} and genomics \cite{WWSHH09,DCApaper,ELLWA13}. These last two fields of research are
particularly active, due to recent advances both in multi-electrode array recording for the former, and in sequencing technology for the latter. These advances provide increasing,  high quality datasets, for hundreds or thousands of neurons or amino-acids / nucleotides. Such large quantities  of data call for statistical modeling,
for example in order to be able to predict the conformation of a protein domain from the sole knowledge of the corresponding genetic sequence 
\cite{proteinfolding_review}. Such modeling can then be tested against the ground truth provided by the well-studied chemistry and biology of neurons and amino-acids, as, for example, X-ray spectroscopy or nuclear magnetic resonance give access to protein conformations, collected into databases such as  the Protein Data Bank (PDB) \cite{PdB}
This approach has been followed very successfully in recent works on
the problem of protein folding \cite{DCApaper}.

The theory of statistical inference has benefited a lot from the contribution of statistical mechanics since the seminal contribution of Jaynes
who introduced the principle of maximum entropy \cite{Ja56}. A more modern point of view that illustrates the introduction of statistical mechanical 
models is given by information geometry \cite{amaribook}, that sees the space of all probability distributions spanning a given statistical model 
as a non-flat manifold, with coordinates given by the parameters of the model. 
This manifold is of course high-dimensional (for discrete datasets like neuronal recordings or genomic sequences), or
infinite dimensional (for datasets taken from continuous variables). The inference can in turn be seen as a an optimization over this 
manifold \cite{Ta00}.

If the dataset is put under a binary form, we can represent it by an $M \times N$ table $\{ \s_i^{(\tau)}\}_{i=1 \cdots N}^{\tau=1 \cdots M}$, 
where $M$ is the number of measurements, and $N$ the number of interacting agents (neurons, amino-acids, traders,...).
A full representation of the data is obtained through its (empirical) moments, the first two being the frequencies $\b{f}^{\rm M}$ and 
pair-wise correlations $\b{p}^{\rm M}$ defined by
\beq
f^{\rm M}_{i} = \frac 1M \sum_{\tau=1}^M \s_i^{(\tau)} \   , \quad  p^{\rm M}_{ij} = \frac 1M \sum_{\tau=1}^M \s_i^{(\tau)} \s_j^{(\tau)} \  ,  
\eeq
and higher-order moments read
\beq
p^{\rm M}_{i_1,\ldots,i_k} = \frac 1M \sum_{\tau=1}^M \s_{i_1}^{(\tau)} \cdots \s_{i_k}^{(\tau)} \  , \forall ~ k = 3 \ldots N  \  .
\eeq
The model parameters can be seen as dual variables, that enforce these specific moments.
The probability distributions that are, by definition, normalized, while having the fixed set of empirical moments $\b{f}^{\rm M}$, 
$\b{p}^{\rm M}, \cdots$ are taken from an exponential family
\beq\begin{split}
& P(\b{\s}) = \frac 1{Z} e^{F(\b{\s})} \  , \\
& F(\b{\s}) \!= \!\sum_{i=1}^N \!h_i \s_i \!+\! \sum_{i<j}\! J_{ij} \s_i \s_j\! +\! \sum_{k=3}^N \sum_{i_1 < \cdots < i_k} \!\!\! J^{(k)}_{i_1 \cdots i_k} \s_{i_1} \cdots \s_{i_k} \ .
\end{split}\eeq
In this dual representation, the model parameters $\bh,\bJ, \cdots$ are fixed by successive Legendre transformations, so that they minimize the
entropy
\beq\begin{split}
\SS[\bh,\bJ;\b{f},\b{p}] = & \ln Z[\bh,\bJ] - \sum_i h_i f_i - \sum_{i<j} J_{ij} p_{ij} \\
& - \sum_{k=3}^N \sum_{i_1 < \cdots <i_k} \left( J_{i_1 \cdots i_k}^{(k)} \right) p_{i_1 \cdots i_k} \  ,
\end{split}\eeq
which, when evaluated at the optimal parameters and at the empirical moments, 
represents the Kullback-Leibler divergence between the empirical distribution 
and the inferred one. Minimizing this Kullback Leibler divergence is clearly equivalent to the maximum likelihood 
estimation of the parameters $\bh,\bJ,\cdots$, while providing a geometric interpretation. 

In real-world datasets, the number of measurements $M$ is not infinitely large compared to $N$, and this renders the estimations
of the empirical moments noisy, and one typically needs a much larger set of measurements to correctly evaluate correlations than to evaluate frequencies. This means in practice that a large number of three- and higher-order correlations are unreliable, 
while even pairwise correlations must be considered as potentially unreliable. Indeed, current challenges in neuroscience and genomics
operate precisely in a regime where $N/M \sim 1$, both $N$ and $M$ are $\OO(10^2)-\OO(10^3)$, and the (connected) correlations 
between the agents $\b{c}^{\rm M} = \b{p}^{\rm M} - {}^t \b{f}^{\rm M} \b{f}^{\rm M}$ are not small. In such situations, 
a sensible choice is to perform the inference on the sub-manifold of distributions that match only the first and second empirical moments. 
Such a limited procedure already gives access to information on the underlying network on which the agents operate, through the 
pairwise couplings $\bJ$. This information can in turn be used to perform, for example, community detection tasks, 
or contact prediction in the case of the protein folding problem \cite{DCApaper}. 

The inference problem is thus ultimately specified by the evaluation of the entropy
\beq
\SS[\b{f},\b{p}] = \underset{\bh}{{\rm inf}} ~ \underset{\bJ}{{\rm inf}}  \left( \ln Z[\bh,\bJ] - \sum_{i=1}^N h_i f_i - \sum_{i<j} J_{ij} p_{ij} \right) \  ,
\label{doubleTL}
\eeq
such that the fields and couplings $\bh^*$ and $\bJ^*$ that solve the inverse problem are given by
\beq
h^*_i = - \frac{\d \SS[\b{f},\b{p}]}{\d f_i} \quad \text{and} \quad J^*_{ij} = - \frac{\d \SS[\b{f},\b{p}]}{\d p_{ij}} \  .
\eeq

A direct method to numerically optimize Eq.~\eqref{doubleTL} reaches unreasonable computation times 
already for $N$$\simeq$$ 20$ \cite{AHS85,Broderick07,SohlDickstein11}. Some speed-up can be obtained by resorting to Newton's method, 
that requires in this case the computation of the Fisher information matrix [18]. See [19] for an application to neuroscience, unfortunately 
limited to the good sampling case for the moment. However the complexity of this method is still exponential, and more advanced methods 
must be used when the number of units is too large, typically larger than a few hundreds.

Another kind of procedure is the Pseudo-likelihood maximization (PLM) \cite{PLoriginal,RW$L_1$0,AE12,Fi14}, which
replaces the standard maximum-likelihood estimator  for the fields and 
couplings by a pseudo-likelihood \cite{MacKay}, i.e. it maximizes
\beq
\la \ln \prod_{i=1}^N P \left( \s_i \left| \{ \s_j \}_{j \ne i} \right. \right) \ra_{\rm M}
\eeq
with respect to the fields and couplings.
In cases where the distribution is that of the Ising model, this leads to finding the optimum for the functional
\beq \begin{split}
\LL_{\rm PL}[\bh,\bJ] = & \sum_i h_i \la \s_i \ra_{\rm M} +2\sum_{i<j} J_{ij} \la \s_i \s_j \ra_{\rm M} \\ 
&  ~ - \sum_i \la \ln \cosh \left( h_i + \sum_{j (\ne i)} J_{ij} \s_j \right) \ra_{\rm M} \   .
\end{split}\eeq
When sampling is large enough ($M \gg N$), the data averages $\la \bullet \ra_{\rm M}$ can be replaced by ensemble 
averages, and $\LL_{\rm PL}[\bh,\bJ] $ is maximized for the same $\bh^*$ and $\bJ^*$ as obtained from the entropy, ensuring that the method is consistent in the limit of large sampling.
However in the regime we are interested in, i.e. $M \sim N$, with both $N$ and $M$ large, and $\b{c}^{\rm M}$ large, the method reaches its limits
\cite{BM09}. Furthermore, the method numerically optimizes over $N+N(N-1)/2$ variables, and thus needs to resort to an uncontrolled number
of iterations for this very large number of unknowns.

Here we focus instead on analytic methods, that will be most suitable in the future for applications to very large datasets.
Our goal is to find approximate functional forms for $\SS[\b{f},\b{p}]$ and obtain an analytical estimate of its minima, in order
to reduce the potentially large (and sometimes uncontrolled) number of iterations that have to be performed with PLM or the Newton's method.
A crude first approximation is the independent model, which, for Ising ($\pm 1$) variables, gives  $Z_{\rm IM}[\b{h}] = \prod_{i=1}^{N} 2 \cosh \left( h_i \right)$, leading to the entropy 
\beq
\SS_{\rm IM}[\b{f}]\! = \! - \! \sum_{i} \! \left[ \! \frac{1-f_i}{2} \! \ln \left(\! \frac{1-f_i}{2} \!\right) \!\!+ \!\!\frac{1+f_i}{2} \! \ln \left(\! \frac{1+f_i}{2} \!\right)\! \right]  \  .
\label{def_SIM}
\eeq
Of course this approximation does not allow to reproduce the strong correlations that are observed in realistic datasets, and one has to go 
further. A possible approach is to perform the two Legendre transforms in Eq.~\eqref{doubleTL} sequentially. The first transform leads us to define
a Gibbs free energy
\beq
\GG[\b{f},\bJ] =  \underset{\bh}{{\rm inf}}  \left( \ln Z[\bh,\bJ] - \sum_{i=1}^N h_i f_i \right)  \  ,
\label{def_GG}
\eeq
from which the entropy is deduced using
\beq
\SS[\b{f},\b{p}] = \underset{\bJ}{{\rm inf}} \left( \GG[\b{f},\bJ] - \sum_{i<j} J_{ij} p_{ij} \right) \  .
\label{second_TL}
\eeq

Approximations for $\GG$ can be obtained for instance through small $\bJ$ expansions, also known as high-temperature expansions \cite{Pl82,GY91}.
The first order in $\bJ$ is the naive mean-field approximation
\beq
\GG^{\rm NMF}[\b{f},\bJ] = \SS_{\rm IM}[\b{f}] + \sum_{i<j} J_{ij} f_i f_j + \OO(\bJ^2) \  .
\label{GG_NMF}
\eeq
Unfortunately, using this approximation, the second Legendre transform cannot be performed since the optimization
over $\bJ$ would give the equation
\beq
p_{ij} = f_i f_j\ ,
\eeq
which has no solution when the data are correlated.
To circumvent this problem, many works have been devoted to the so-called linear response method \cite{KR98}.
Instead of performing the second transformation one takes advantage of the exact relation
\beq
\chi^{-1}_{ij} = - \frac{\d^2 \GG[\b{f},\bJ]}{\d f_i \d f_j} \  ,
\label{linear_response}
\eeq
where $\b{\chi}$ is the connected correlation function of the model at fixed frequencies $\b{f}$ and couplings $\bJ$.
The inference for the couplings is done in that case by searching for the $\bJ^*$ that satisfy 
Eq.~\eqref{linear_response}
with the exact correlation function $\b{\chi}$ replaced by the correlation function measured in the data 
$\b{\chi} \to \b{c}^{\rm M}$. When the lowest order approximation in Eq.~\eqref{GG_NMF} is used, this procedure is termed
naive mean-field inference (NMF), or Direct Coupling Approximation (DCA)  for its generalization to non-binary variables \cite{DCApaper}.
Going to second order in $\bJ$ in the expansion of $\GG$ leads to the so-called Thouless-Anderson-Palmer (TAP) procedure, used for 
example for machine learning in \cite{GTK15}. The corresponding Gibbs free energy reads
\beq
\GG[\b{f}\!,\!\bJ]\!=\! \GG^{\rm NMF}[\b{f}\!,\!\bJ] + \frac{1}{2}\!\! \sum_{i<j} L_{ii} J_{ij}^2 L_{jj} + \OO(\bJ^3) \  ,
\label{TAP}
\eeq
where $\b{L}$ is the (diagonal) matrix  of self-correlation of independent variables obtained through
\beq
\left( \bL \right)^{-1}_{ij} = - \frac{\d^2 \SS_{\rm IM}[\b{f}]}{\d f_i \d f_j} = \frac{1}{1-f_i^2} \d_{ij} \ .
\eeq
Whatever order in the small $\bJ$ expansion is used, resorting to linear response leads to incoherences
because the diagonal part of Eq.~\eqref{linear_response} cannot be satisfied properly. 
A host of works have been devoted 
to correct with ad-hoc methods this consistency problem, that all relate to the so-called adaptive TAP approach of
Opper and Winther \cite{OW01}, and that are usually termed diagonal matching methods
\cite{Ta98,HK13,Ki14,RR13a,RR13b,Yasuda2013}. 

The next logical step, following the tradition of theoretical physics, 
is to use further diagrammatic resummations. Resumming two-spin diagrams in the 
Gibbs free energy leads to the Bethe approximation.
To obtain the Bethe free energy, one calculates the contribution to $\GG$ of all pairs of variables $(i,j)$ interacting independently from the 
other pairs. One obtains the result
\beq
\GG[\b{f},\bJ] \approx \SS_{\rm IM}[\b{f}] + \sum_{i<j} \D \GG_{ij}[\b{f},\bJ]  \  ,
\label{Bethe}
\eeq
where $\D \GG_{ij}$ is the difference between the free energy of the isolated pair $(i,j)$ (interacting through the coupling $J_{ij}$), 
and the free energy of two independent variables $i$ and $j$.  Explicit formulas are cumbersome and can be found in 
\cite{WT03,Ricci-Tersenghi2012}. The Bethe approximation is often solved by message-passing algorithms 
\cite{MM09,NB12a,Ricci-Tersenghi2012,RR13a,RR13b}. Unfortunately, these analytical methods are 
generically unable to infer correctly inside a low temperature phase, when correlations are strong, or sampling is low, 
see for example \cite{SM09,Ricci-Tersenghi2012}. In addition, the Bethe approximation is exact on trees, whereas
strongly interacting units define (by definition) very densely connected interaction graphs that contain many loops \cite{MR05,NB12a}. 

Since the correct procedure is to perform the second Legendre transform with respect to $\bJ$, the natural step
is to use the small $\bJ$ expansion in Eq.~\eqref{TAP} and turn it in an expansion of the entropy in powers of connected correlations.
This is the small-correlation expansion of Sessak and Monasson (SM) developed in \cite{SM09}.
We define an off-diagonal correlation matrix $\bt{c} \equiv \b{c} -\b{L}$, 
and the small-correlation expansion is an expansion in powers of $\bt{c}$, the first term of which is
easily deduced from the TAP free-energy in Eq.~\eqref{TAP} to find
\beq
\SS[\b{f},\b{p}] = \SS_{\rm IM}[\b{f}] - \frac{1}{2} \sum_{i<j} \frac{\t{c}_{ij}^2}{L_{ii}L_{jj}} + \OO(\bt{c}^3) \  .
\label{small-c_lowest}
\eeq
Higher-order terms can be calculated, however since the correlations in realistic data are large \cite{SM09}
the obtained series is divergent, and resummations must be used. The natural thing to do is to resum ring diagrams, 
which leads to the approximation
\beq
\SS^{\rm ring}[\b{f},\b{p}] \approx \SS_{\rm IM}[\b{f}] + \frac 12 \Tr \left[ \ln \bc  - \ln \b{L} \right] \  .
\label{ring}
\eeq
However, this method, even coupled to two-spin and three-spin resummation was found to be extremely sensitive to 
sampling noise \cite{SM09}, rendering it impractical.  

Finally, another alternative to find a theoretically well-founded approximation for $\SS$ it the adaptive cluster expansion (ACE) of \cite{CM11,CM12}. 
One expands the partition function in the equivalent for spin systems
of the virial coefficients, and gradually incorporates more and more diagrams depending on their 
information content, measured through their contribution to the entropy $\SS$. The starting point of the expansion
is the independent spin model, and one can then incorporate the interactions between units by 
considering again the pairs of spins as independent, which gives an approximation for the entropy 
\beq
\SS[\b{f},\b{p}] \approx \SS_{\rm IM}[\b{f}] + \sum_{{\rm pairs }~ i,j} \D \SS^{(2)}_{ij}[\b{f},\b{p}] \  ,
\eeq
similarly to the two-spin diagrams resummation for $\GG$ described above. 
However, in the case of ACE, the summation runs over a given set of pairs $i,j$, that must be chosen beforehand. 
This procedure can be continued by taking into account larger and larger subsets of spins (called ``clusters'' in that context) into
account. The adaptive cluster expansion selects relevant clusters of spins depending on their final contribution to
the entropy $\SS$. However when clusters are too large (already for triplets of spins in the case of $q>2$), it is too
cumbersome to perform the double Legendre transform analytically, and the algorithm selects a trial set of clusters
in the expansion of $Z[\bh,\bJ]$, and optimize numerically over $\bh$ and $\bJ$ to compute the entropy $\SS$. 
The algorithm is very efficient in avoiding oversampling, and optimal either when sampling noise is large, or when 
the interaction graph of the units is  sparse enough: in both cases, only small clusters of spins will be selected 
(in these cases the algorithm was shown 
 to be able to saturate the Cram\'er-Rao bound for the variance of the maximum likelihood estimator \cite{CM11}). 
However 
the complexity of the algorithm is exponential in the size of the clusters that have to be taken into account. 
When the interaction graph is dense and the correlations are large, 
and when the number of states $q$ is large, for example for protein data, 
or when the number of units $N$ is large, the algorithm hits its limits. 
For example $N=27$ strongly coupled amino-acids (i.e. with $q=20$ or $21$) 
in a lattice model of protein already pushes the algorithm to its
limits, see \cite{hugomonasson} for an example.  \\

One of the main source of difficulty in the inference problem is the presence of sampling noise. 
Whereas the functional $\ln Z[\bh,\bJ]-\sum_i h_i f_i - \sum_{i<j} J_{ij} p_{ij}$ is always a strictly convex function of $\bh,\bJ$, whatever the values of $\b{f}$ and $\b{p}$ are, it is not guaranteed to have its minimum at a finite value of the fields and couplings. 
To bypass this limitation, one can simply add a regularization term, which has the Bayesian
interpretation of adding a prior to the parameters $\bh$ and $\bJ$. With the addition of the regularization term the posterior probability 
can be maximized instead of the likelihood, which leads to considering a modified entropy functional
\beq
\SS_{\rm reg}[\b{f},\b{p}]  = \underset{\bh,\bJ}{{\rm inf}} \left( 
\begin{array}{ll}
& \displaystyle \ln Z[\bh,\bJ] - \sum_i f_i h_i \\
& \\
& \displaystyle - \sum_{i<j} p_{ij} J_{ij} - \frac 1M \ln P_{\rm prior}[\bh,\bJ] 
\end{array}
\right)
\eeq
where $P_{\rm prior}[\bh,\bJ]$ is the prior probability on $\bh$ and $\bJ$.
One can consider for example the class of $L_{n}$ regularization on the couplings only, 
which leads to minimizing over $\bJ$ (since the fields are not regularized at all) the regularized Gibbs free-energy
\beq
\GG_{\rm reg}[\b{f},\bJ] = \GG[\b{f},\bJ] + \frac {\eta}{n M} \left| \left| \bJ \right| \right|_{n}^n
\label{GG_reg}
\eeq
where $\left| \left| \bullet \right| \right|_{n}$ is the $L_n$ norm, and $\eta$ is the strength of the regularization. 
The cases $n=1$ or $2$ are very popular in the statistics community since the
former selects sparse models and the latter selects models with small parameters. With this addition, the inference problem has now a unique and finite 
solution. For these reasons, the PLM and ACE algorithms need such a regularization to not be trapped in locally flat directions during their 
numerical optimizations over $\b{p}$. As far as analytical schemes are concerned, we see that the regularization term must, by construction, be 
proportional to the inverse number of samples, i.e. should be small when sampling is large. 
In principle, this issue can thus be taken into account perturbatively once the perfect sampling problem
has been tackled. 

The paradigmatic illustration of the necessity of regularization is given by the NMF inference,
that amounts to invert the data correlation matrix, as can be seen by solving the inference problem starting from the ring entropy
shown in Eq.~\eqref{ring}, or indifferently from the NMF+linear response scheme:
\beq
J_{ij}^{\rm NMF} = - \left. \frac{\d \SS^{\rm ring}[\b{f},\b{p}]}{\d p_{ij}} \right|_{\b{p}^{\rm M}} = - \left( \b{c}^{\rm M} \right)^{-1}_{ij} \  .
\label{eq_Jring}
\eeq
The empirical correlation matrix $\b{c}^{\rm M}$ is usually rank-deficient in realistic datasets, 
preventing NMF to be applicable as is. The fact that the correlation matrix is not invertible corresponds to a situation where
the functional $\ln Z[\bh,\bJ] - \sum_i h_i f_i^{\rm M} - \sum_{i<j} J_{ij} p_{ij}^{\rm M}$ is minimized by infinite values of some of the
couplings or fields. However, in effect the problem is that the ring entropy in Eq.~\eqref{ring} is not differentiable at $\b{f}^{\rm M},\b{p}^{\rm M}$.
We see that in that case two issues get mixed: the non-existence of the solution to the unregularized inference problem and the fact that the 
ring entropy is ill-behaved. Adding a regularization term as discussed above restores the ability of NMF to infer coupling parameters,
see \cite{Jones12,Andreatta13} for examples of applications to realistic data. 
Another possibility is to add pseudocounts to the data before computing the one- and two-point marginals, that is, in the case 
of binary variables perform the modifications \cite{BCLM14}
\beq\begin{split}
& f_i^{\rm M} ~ \to ~  (1-\a) f_i^{\rm M} \quad \forall ~ i \ ,\\
& p_{ij}^{\rm M} ~ \to ~  (1-\a) p_{ij}^{\rm M} + \a ~ \d_{ij} \quad \forall ~ i,j \ .
\end{split}\eeq
The same kind of modification can be applied for multi-index variables, see \cite{DCApaper} for protein data, 
which amounts to add a prior to unobserved data \cite{gaussian}. 
Both types of regularization allow to compensate for the rank 
deficiency of $\b{c}^{\rm M}$, and restore the ability of the mean-field inference to infer couplings. 
Interestingly, it was found that for NMF, large regularizations $\eta$ (of the order of $M$) must be chosen 
to have a quantitative result, which is not the case for ACE, where the regularization can (and should) be chosen of the order one.
This particular feature points toward a pathology in the analytical formulations at hand, that is not present in the more direct, 
methods like ACE and PLM, that numerically optimize over $\bJ$. 

In this paper, we continue the procedure of diagrammatic resummations for the entropy functional, initiated in \cite{SM09},
in order to obtain an inference procedure that is stable with respect to sampling noise.
Instead of relying on a small correlation expansion, and inspired by a field theoretic point of view, we set up a ``loop expansion'' of the entropy 
(not to be confused with the loops of an interaction graph), and show that 
it contains and generalize the majority of analytical methods that are based on mean-field methods or high 
temperature / small correlation expansions. Our procedure is shown to resum a large number of diagrams in the 
small correlation expansion of SM, including those leading to NMF inference,  and diagonal 
matching methods, thus providing a unifying picture for all these works, along with
an inference algorithm that is able to infer quantitatively fields and couplings across the whole phase diagram of 
spin glasses, without being critically sensible to sampling noise as in \cite{SM09}. Despite introducing an iterative scheme to solve for 
$N$ auxiliary variables, the complexity of the algorithm is still very low, since it requires only matrix inversions and multiplications. 
We demonstrate that even deep in the spin 
glass regime, and in the presence sampling noise, our inference procedure still produces meaningful results, 
whereas other analytical methods fail badly when they are not regularized. 
A key feature of our approach is that it does not rely on inversion of the correlation matrix, 
thanks to the introduction of $N$ auxiliary variables that are fixed solely by the data itself, 
so that the method works even when $\b{c}^M$ is rank deficient, without relying on any pseudocount or $L_1$ or $L_2$ penalties. 
Indeed we argue that the class of diagrams that we resum give large contributions in the under sampling regime $N \sim M$, so that their
omission in previous works led to inference schemes that were unstable when sampling is poor.

The plan of the paper is the following: in Section \ref{sec_RMF}, we  derive in details our main results given in 
Eqs.~(\ref{def_D})-(\ref{RMF_couplings}) and (\ref{RMF_fields})-(\ref{result}). In Section \ref{sec_exp}, we make the connexion with the well-known 
high-temperature and small-correlation expansions, as well as the diagonal matching tricks. The reader
uninterested by technical details can safely skip to Section \ref{sec_SK}, where we test our results on the Sherrington-Kirkpatrick model. 
We show that our method (without regularization) outperforms the other analytical inference methods, even when they are regularized. 
We show that on top of being stable across the whole phase diagram, our method provides meaningful fields and couplings, also at the level
of individual probabilities of configurations, a feature that is inaccessible to most of the other methods, and poorly performed by NMF combined 
with regularization and diagonal matching, or by the Gaussian model.
In Section \ref{sec_potts}, we adapt the expansion to Potts variables, and in Section \ref{sec_2ndorder}, we push the expansion to the next order. 
Finally we give our conclusions in Section \ref{sec_concl}.

\section{Resummed mean-field approximation \label{sec_RMF}} 

We seek for an approximation of $\SS$ based (using Eq.~\eqref{second_TL}) on an expansion of $\GG$ in terms of a small parameter, while improving upon previously known approximations of the Gibbs free energy, such as the high-temperature expansion. For that purpose, we use the following exact equation for the Gibbs free energy, known as the Wetterich equation in the context of quantum field theory \cite{We93},
\beq \begin{split}
& \partial_\be \GG[\b{f},\be\bJ] =\sum_{i<j} J_{ij} \ch_{ij}(\be) + \sum_{i<j} J_{ij} f_i f_j \ ,\\
& \text{where} \quad \left( \b{\chi}(\be)^{-1} \right)_{ij} = - \frac{\d^2 \GG[\b{f},\be \bJ]}{\d f_i \d f_j}\ .
\end{split}
\label{Wetterich}
\eeq
This equation has been the starting point to non-perturbatively tackle a variety of problems stemming from quantum and 
statistical field theories \cite{Berges2002}, with recent development for classical and quantum lattice models \cite{Machado2010,Rancon2011,Rancon2011a}. 
Here the function $\bch$ is the exact correlation 
function of the Ising model, and thus solving Eq.~\eqref{eq_integralWett}  exactly is as hard as computing the partition function directly, and one has to 
resorts to approximate solutions. One could expand in powers of $\bJ$ to straight-forwardly recover the high-temperature expansion of 
\cite{Pl82,GY91}, followed by a Legendre transform with respect to $\bJ$ to recover the results of \cite{SM09}, see Section \ref{sec_exp} and 
Appendix \ref{app_exp} for details.

Instead, we first formally integrate Eq.~\eqref{Wetterich} and obtain (see also \cite{OW01}),
\beq\begin{split}
& \GG[\b{f},\bJ] = \SS^{\rm IM}[\b{f}] +  \sum_{i<j} f_i J_{ij} f_j +  \int_{0}^{1} d\be ~ \sum_{i<j} J_{ij} \chi_{ij}(\be)  \  .
\end{split}
\label{eq_integralWett}
\eeq
We see that under this integral form, the naive mean-field result is obtained when the integral is dropped. 
Starting from this observation, we build a functional $\GG_{\e}$, that interpolates between the mean-field result and the 
exact one, defined by the solution to the system
\beq\begin{split}
& \GG_{\e}[\b{f},\bJ] = \GG^{\rm NMF}[\b{f},\bJ] +  \e \int_{0}^{1} d \be  ~ \sum_{i<j} J_{ij} \chi_{ij}(\be,\e)  \  , \\
& \text{where} \quad \left( \b{\chi}(\be,\e)^{-1} \right)_{ij} \equiv - \frac{\d^2 \GG_{\e}[\b{f},\be \bJ]}{\d f_i \d f_j} \  .
\end{split}\eeq
This particular choice of dependencies on $\e$ ensure that the $\e$$\to$$0$ limit recovers the NMF result, while the 
$\e$$\to$$1$ limit recovers the full theory. We then construct a corresponding interpolating entropy $\SS_\e$ by setting
\beq
\SS_{\e}[\b{f},\b{p}] = \underset{\bJ}{{\rm inf}} \left( \GG_{\e}[\b{f},\bJ] - \sum_{i<j} J_{ij} \left[ f_i f_j + \e (p_{ij}-f_i f_j) \right] \right) \ ,
\eeq
implying $\SS_{\e=0}=\SS^{\rm IM}$ and $\SS_{\e=1}=\SS$. Our procedure thus constructs
a new kind of expansion of the entropy starting from the independent model approximation (and correspondingly, an expansion of the Gibbs free energy starting from NMF).

We now assume that $\GG_{\e}$ and $\SS_\e$ are analytic functions of $\e$, and seek for their series expansions in powers of $\e$, which read
\beq
\GG_\e = \sum_{k=0}^{+\io} \e^k\GG^{(k)}  \  , \quad\quad \SS_{\e} = \sum_{k=0}^{+\io}\e^k \SS^{(k)}  \  .
\eeq
These expansions are the counterpart, for spin systems, of the well-known loop expansions originated from quantum field theory \cite{zinn,Litim2002}. 
We show in Appendix \ref{app_saddle} another formulation of this expansion in terms of a saddle-point evaluation of a certain path integral, that 
makes this connection clearer. However, the point of view we adopt is by far more effective, when actual calculations are concerned, 
than the field theoretic one. 

Starting from the expansion to order one in $\e$ of $\GG_\e$, we deduce the expansion of $\bch(\be,\e)$ in powers of $\e$, leading to 
\beq
\bch(\be,\e) = \bch^{(0)}(\be) + \e \,\bch^{(0)}(\be) \frac{\d^2 \GG^{(1)}[\b{f},\be \bJ]}{\d \b{f}\d \b{f}} \bch^{(0)}(\be) + \OO(\e^2)\       ,
\label{exp_c}
\eeq
where
\beq
\begin{split}
(\bch^{(0)}(\be))^{-1}_{ij} &=-  \frac{\d^2 \GG^{\rm NMF}[\b{f},\be \bJ]}{\d f_i \d f_j}=  L^{-1}_{ij} - \be J_{ij} \ .
\end{split}
\label{eq_cNMF}
\eeq
Inserting these results in Eq.~\eqref{eq_integralWett}, we obtain after a trivial integration the first order result
\beq
\GG^{(1)}[\b{f},\bJ] = - \frac 12 \Tr \left[ \ln \left( \bL^{-1}- \bJ \right) - \ln \bL^{-1} \right]\ .
\eeq
Our first order (in $\e$) approximation for the Gibbs free energy is thus 
\beq\begin{split}
\GG_\e[\b{f},\bJ] &=  \SS_{\rm IM}[\b{f}] +  \sum_{i<j} J_{ij} f_i f_j  \\
& - \frac \e 2 \Tr \left[  \ln \left( \bL^{-1} - \bJ \right) - \ln \bL^{-1} \right] + \OO(\e^2)\ .
\label{eq_GG1loop}
\end{split} \eeq
The net effect of our first order procedure is to resum ring diagrams in the Gibbs free energy. 
This could have been done by hand 
simply by looking at the series expansion given by the high-temperature expansion, and was indeed shown in \cite{CM12}.
However we will show in Section \ref{sec_potts} 
that our method gives a systematic procedure, that could not be performed by hand beyond the first order. 

Starting from this improved Gibbs free energy, we evaluate the first order contribution to the entropy. 
The optimal couplings $\bJ^*$, that minimize $\GG$, are obtained, to first order in $\e$, by solving the equation
\beq \begin{split}
c_{ij} \equiv p_{ij} - f_i f_j = \left( \bL^{-1} -  \bJ^* \right)^{-1}_{ij} + \OO(\e) \quad \forall ~ i<j \ ,
\end{split} 
\label{eq_cJ}
\eeq

Keeping in mind that by definition $\bJ^*$ is a symmetric matrix with zeros on the diagonal and with $J^*_{ij}$ for $i<j$ outside the diagonal,
Eq.~\eqref{eq_cJ} therefore gives $N(N-1)/2$ non-linear equations for the $N(N-1)/2$ unknowns $J^*_{ij}$, which could be solved for example 
numerically. It is crucial to keep in mind that Eq.~\eqref{eq_cJ} cannot be inverted in a matrix sense because we do not have an equation 
for $i$$=$$j$. A naive inversion would lead to the ring entropy discussed in Sec.~\ref{sec_exp}, and recovers the NMF+linear response result for 
the couplings. We assume now that a solution has been found for this system. In that case, we can define a diagonal matrix 
$\b{D}$ (which depends on $\b{f}$ and $\b{p}$) by
\begin{equation}
D_{ii} = \left( \bL^{-1} - \bJ^* \right)^{-1}_{ii} - c_{ii} \quad \forall ~ i  \ ,
\end{equation}
We have now the property, valid in the matrix sense, that
\beq
\bL^{-1} - \bJ^* = \left( \b{c} + \b{D} \right)^{-1}\ .
\label{Jstar_matrix}
\eeq
which allows us to find a posteriori  the explicit value of $\bJ^*$ and $\b{D}$ by evaluating on and out of the diagonal.
We find, on the diagonal, a set of $N$ equations that solve the $N$ unknown elements of $\b{D}$ (importantly, independent of $\bJ^*$),
\beq
\left( \b{c} + \b{D} \right)^{-1}_{ii} = L^{-1}_{ii} \quad \forall ~ i\ .
\label{def_D}
\eeq
A numerical procedure to compute $\b{D}$ is discussed in Appendix \ref{app_numD}. 
We finally get the equation for the inferred couplings by evaluating Eq.~\eqref{Jstar_matrix} outside the diagonal
\beq
J^*_{ij} = - \left( \b{c} + \b{D} \right)^{-1}_{ij} + \OO(\e) \quad \forall ~ i<j\ .
\label{RMF_couplings}
\eeq
Note that the matrix $\bD$ is a tool to formally invert Eq.~\eqref{eq_cJ}, and is thus very different from the diagonal-
matching tricks sometimes used to solve the inconsistency of the NMF or TAP approximations \cite{Ki14}.  
A direct comparison between the small correlations expansions of both 
\eqref{eq_cJ} and \eqref{RMF_couplings} readily confirms that they both contain the same diagrams, see the 
discussion in the next section.

The equation for the fields is easily deduced from the relations
\beq
h^*_i = - \left.\frac{\d \SS_{\e}}{\d f_i}\right|_{\b{p}} = - \left. \frac{\d \GG_{\e}}{\d f_i} \right|_{\bJ^*} \ ,
\eeq
and we find
\beq
h_i^* = \tanh^{-1}(f_i) - \sum_{j (\ne i)} J_{ij}^* f_j + \e \frac{D_i f_i}{(1-f_i^2)^2}  + \OO(\e^2) \ .
\label{RMF_fields}
\eeq
From the approximation for the couplings in Eq.~\eqref{RMF_couplings} we obtain our resummed mean field (RMF) approximation for the cross-entropy
\beq\begin{split}
\SS^{\rm RMF}[\b{f},\b{p}] = & \SS_{\rm IM}[\b{f}]  - \frac \e2 \Tr \left(\b{D}\bL^{-1}\right)   \\
& + \frac \e 2 \Tr \left( \frac{}{}\!\! \ln \left( \b{c} + \b{D} \right) - \ln \bL \right) + \OO(\e^2)\ .
\end{split}
\label{result}
\eeq
The calculation can be easily 
continued to second order in $\e$ for Potts or Ising variables, and to third order for Ising variables, 
although it becomes gradually more tedious, the number   terms increasing rapidly (see Sec.~\ref{sec_potts}).

\section{Connections with previous analytical approaches \label{sec_exp}}

We now discuss the connections between our RMF approximation and previous works, such as the small correlation expansion of SM, and  clarify the effect of the matrix $\bD$ introduced to invert Eq.~\eqref{eq_cJ}.
First of all, let us note that one can recover the Plefka expansion (high temperature, i.e. small $\bJ$, expansion) of $\GG[\b{f},\bJ]$ up to arbitrary 
order starting from the Wetterich equation, as shown in Appendix \ref{app_exp}. Of course, the expansion in $\e$ resums an infinite number of 
terms in power of $\bJ$, and is thus much more powerful.  
For example, at order $\e$, Eq.~\eqref{eq_GG1loop} expanded to order $\bJ^2$ gives back the TAP result of Eq.~\eqref{TAP} 
(after setting $\e=1$).

We have checked explicitly that the expansion of $\GG_{\e}$ to order $\e^3$ indeed contains all terms of its expansion to 
order $\bJ^4$. 
In addition, 
we show formally in Appendix \ref{app_proof} that the $\e$ expansion of $\GG_{\e}$ at order $\e^n$ contains all terms of the 
Plefka expansion to order $\bJ^{n+1}$. The fact that $\GG_{\e}$ at order $\e^n$ is exact to order $\bJ^{n+1}$ obviously transfers to the expansion 
of $\SS_{\e}$, and this proves that our approximation scheme contains the small correlation expansion while resumming a further (infinite)
class of diagrams at each order in $\e$.

Starting from Eq.~\eqref{result}, the small correlation expansion of $\SS^{\rm RMF}$  reproduces the expansion of \cite{SM09}, 
for example at the lowest order one finds back the lowest order in the small correlation expansion shown in Eq.~\eqref{small-c_lowest}. 
In \cite{SM09}, the authors devised resummations of the small correlation expansion. 
In particular,  they resum an infinite series of terms for the couplings  ($i\neq j$) that corresponds to the ring entropy given in 
Eq.~\eqref{ring}. These diagrams correspond to the diagrams shown in the first line in Fig.~\ref{fig_diagJ}.

Note that the optimization equation for the couplings derived from $\SS^{\rm ring}$
corresponds to the NMF result $\bJ^{\rm NMF}=-(\bc^{\rm M})^{-1}$ usually obtained \emph{using linear response}, 
a method that is not consistent since we have in that case
\beq
- \frac{\d^2\GG^{\rm NMF}}{\d f_i \d f_i}=\frac 1{1-f_i^2} \neq (\bc^{-1})_{ii} \  ,
\eeq 
and this equation is not satisfied for $i=j$.
But once more, we stress that the Legendre transform of $\GG^{\rm NMF}$ with respect to $J_{ij}$ is not well defined 
and thus $\SS^{\rm NMF}$ does not exist strictly speaking, although $\SS^{\rm ring}$ would be the closest, and most natural, proxy for it.
This feature explains why the linear response method is successful in general: it allows one to resum a certain class
of higher-order diagrams of the second Legendre transform, without having to explicitly perform the Legendre 
transformation, as it  is indeed well known in statistical field theory \cite{zinn}. 
This, however, comes at the cost of inconsistencies on the diagonal part of the inverse correlation matrix.

\begin{figure}[t]
\center
\includegraphics[width=8cm]{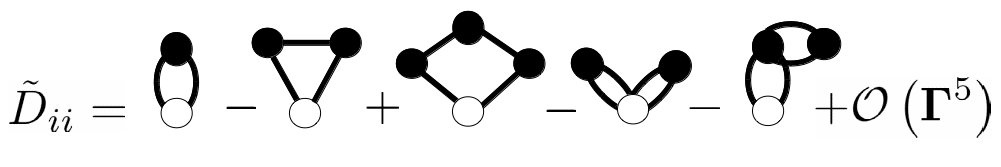}
\caption{Diagrammatic expression of the small-correlation expansion of the rescaled $\bt{D}$ matrix. 
Open circles represent the site $i$, the lines lines are $\b{\G}$ factors and filled circles mean summation over a site index.}
\label{fig_diagsD}
\end{figure}

A very interesting fact, to our knowledge not yet discussed in the literature, is that the fields obtained from $\SS^{\rm ring}$ correspond to the fields obtained from the NMF with the diagonal matching trick (which has been shown to be equivalent to the adaptive TAP method in the direct problem \cite{Yasuda2013}). Indeed, we find
\beq
h^{\rm ring}_i=\tanh^{-1}(f_i)- \sum_{j(\neq i)}J^{\rm ring}_{ij}f_j+\left(\left(\bc^{-1}\right)_{ii}-L_{ii}^{-1}\right)f_i\ ,
\label{eq_hring}
\eeq
which is the same result than obtained from the NMF diagonal matching free energy 
\beq
\GG^{\rm DM}[\b{f},\bJ]=\GG^{\rm NMF}[\b{f},\bJ]+\frac12\sum_i \Lambda_i (1-f_i^2) \  , 
\eeq
using linear response \cite{Ki14}, that we recall now for the sake of completeness. 
Linear response for $\GG^{\rm DM}$ gives
\beq
\begin{split}
h_i=\tanh^{-1}(f_i)- \sum_{j(\neq i)}J_{ij}f_j+\Lambda_i f_i \ , \\
(\bc^{-1})_{ij}=-J_{ij}+\d_{ij}\left(\frac{1}{1-f_i^2}+\Lambda_i\right)\ ,
\end{split}
\label{eq_linDM}
\eeq
which yields, when solved for $h_i$ and $J_{ij}$,  Eqs.~\eqref{eq_hring} and  \eqref{eq_Jring}. 
Therefore, using $\SS^{\rm ring}$ gives a more rigorous way to derive the NMF inference than linear response of 
adaptive TAP, or diagonal matching trick,  as it is based on the proper object to perform the inference, the entropy, 
which in that case corresponds to the resummation of all ring diagrams. This is also an a posteriori justification of the 
improvement of the diagonal trick upon the more ``naive'' NMF fields (corresponding to Eq.~\eqref{eq_hring} without 
the last term), since $h_i^{\rm ring}$ corresponds to a more consistent inference based on $\SS^{\rm ring}$.

\begin{figure}[t]
\center\includegraphics[width=8cm]{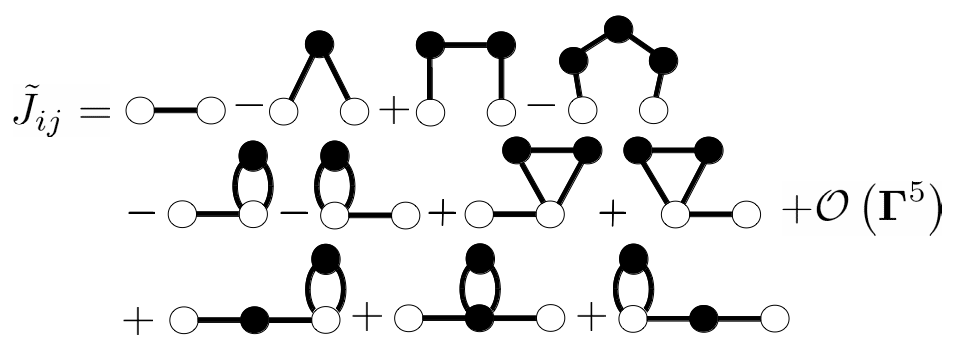}
\caption{Diagrammatic expression of the small-correlation expansion of the couplings obtained through RMF. Same notations as in Fig.~\ref{fig_diagsD}, with one open circle representing the site $i$ and the other the site $j$. The first line corresponds to the truncated expansion of NMF couplings to order $\b{\Gamma}^4$ (ring diagrams), the other terms coming from that of $\b{D}$.}
\label{fig_diagJ}
\end{figure}

We want now to compare the RMF entropy with the ring entropy, which contains both NMF and the diagonal matching 
method for the fields. To do so, we need to expand the matrix $\bD$ in power of the off-diagonal part of the correlation matrix $\b{c}$.
We call $\bt{c}$ the matrix with elements $c_{ij}$, and zeros on the diagonal.
To alleviate the notations, we define a rescaled correlation matrix $\b{\G}$ 
\beq
\b{\G} \equiv \bL^{-1/2}  \bt{c} ~ \bL^{-1/2} \ ,
\label{def_Gamma}
\eeq
(note that $\b{\G}$ inherits from $\bt{c}$ the property that it has zeros on its diagonal),
and the small correlation expansion amounts to an expansion in powers of $\b{\G}$. 
We define a rescaled $\bt{D}$ matrix by
\beq
\bt{D} \equiv \bL^{-1/2} \b{D} \bL^{-1/2} \  ,
\label{D_rescaled}
\eeq
and Eq.~\eqref{def_D} then becomes
\beq
\left( \mathbb{1} +  \b{\G} + \bt{D} \right)^{-1}_{ii} = 1 \quad \forall ~ i \ , 
\eeq
which is easily expanded in powers of $\b{\G}$ to find
\beq
\begin{split}
\bt{D}_{ii}&= \left(\b{\G}^2\right)_{ii} - \left(\b{\G}^3\right)_{ii} \\
& + \left( \left(\b{\G}^4\right)_{ii} -\left(\b{\G}^2\right)_{ii}^2-\sum_{j,k}\G_{ij}^2\G_{jk}^2\right)+\OO(\b{\G}^5) \ .
\end{split}
\eeq
The diagrammatic representation of this expansion is shown in Fig.~\ref{fig_diagsD} up to order 4 in $\b{\G}$, and this 
shows that $\bD$ resums rings of $\t{\bc}$ going from the site $i$ through an arbitrary number of intermediary sites before going back to $i$.
From the expansion of $\b{D}$, we deduce the expansion of rescaled couplings
\beq
\t{\bJ} \equiv \bL^{1/2}\bJ^*\bL^{1/2} \  ,
\eeq
that is shown diagrammatically in Fig.~\ref{fig_diagJ}.
The first line of the expansion corresponds to the (truncated) sum of ring diagrams corresponding to the NMF 
couplings, whereas the other terms come from the expansion of the $\bD$ matrix. 
We also see that $\SS^{\rm ring}$ is recovered, and thus NMF and the diagonal matching trick, if one takes $\bD=0$. Indeed we have, 
after setting $\e=1$, the relation
\beq
\SS^{\rm RMF} = \SS^{\rm ring} + \frac 1 2 \Tr \left[  \ln \left( \mathbb{1} + \b{c}^{-1} \b{D} \right) - \bL^{-1} \b{D} \right] \  ,
\eeq
which demonstrates that our framework indeed goes beyond these previous methods in terms of diagram 
resummations. A final remark is that, as already stated above, the introduction of the $\b{D}$ matrix was done only in order 
to push the analytical calculations further, and it is easily verified that the small correlation expansion of 
Eq.~\eqref{eq_cJ} coincides with the expansion of Eq.~\eqref{RMF_couplings} combined with the expansion of 
$\b{D}$, i.e. to the diagrams of Fig.~\ref{fig_diagJ}. The introduction of $\b{D}$ is thus in no way necessary, and does not correspond in any way to some kind 
of diagonal matching method.

\begin{figure}[t!]
\includegraphics[width=4cm]{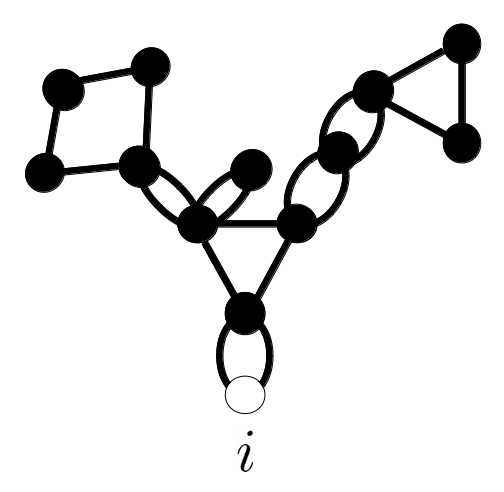}
\caption{A typical diagram contributing to $\bt{D}$, same conventions than in Fig.(\ref{fig_diagsD}).}
\label{fig_big_diagram_D}
\end{figure}

We have shown the small-correlation expansion of RMF, but of course  the couplings in Eq.~\eqref{RMF_couplings}
contains an infinite number of diagrams beyond those shown in Fig.~\ref{fig_diagJ}. A typical diagram contributing to $\b{D}$, that is resummed 
by our approximation is shown in Fig.~\ref{fig_big_diagram_D}. All such ``cactus diagrams" \cite{PP95} are resummed in $\b{D}$, and the 
usual ring diagrams contributing to the NMF result must be dressed at each vertex by all such diagrams to obtain the RMF couplings 
(see Fig.~\ref{fig_diagJ} for an illustration: the single $\b{\G}$ link on the first line is dressed by the first two contributions to $\b{D}$ 
to give the diagrams ont the second line).

This clearly shows that our approximation resums a whole class of `cactus' diagrams  in addition to the 
simple ring diagrams that lead to NMF. This resummation would be very hard to guess simply upon looking at the 
diagrams of the small-correlation expansion
of \cite{SM09}, which explains why it has been over-looked. 

Finally, in \cite{SM09}, the authors combine their ring resummation with the resummation of all two-spin diagrams
(and even three-spin diagrams in the $\b{f}=0$ case), and we show in Appendix \ref{app_2spin} that the same procedure can 
be applied in our framework, since one simply has to identify the contribution of two- or three-spin diagrams
in the RMF result.  We leave the issue of testing the RMF plus two-spin inference procedure for future work, and focus on the simpler RMF inference in the following.

\section{Tests on the Sherrington-Kirkpatrick model \label{sec_SK}}

\begin{figure}[t!]
\center\includegraphics[width=8cm]{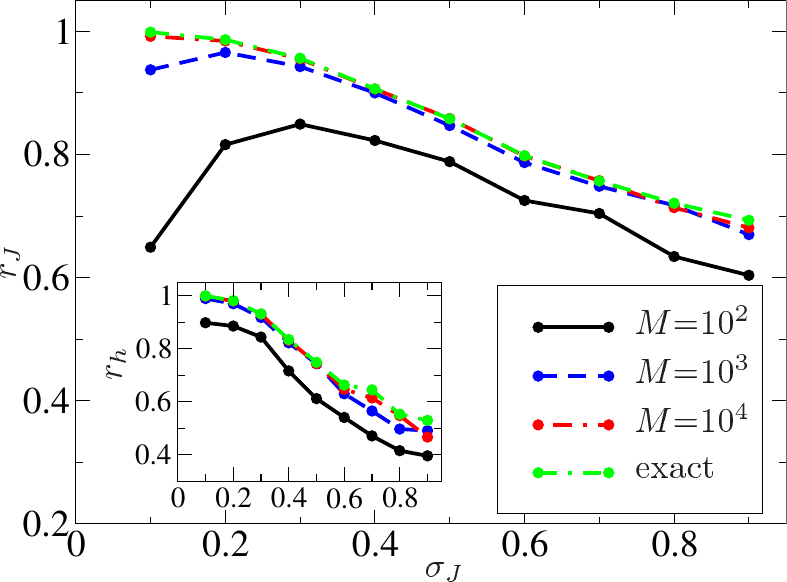}
\caption{Average Pearson correlation $r_J$ (see Eq.~\eqref{def_pearson} and its discussion) between the true couplings and the couplings inferred from RMF for various sampling $M=10^2-10^4$, as well as in the perfect sampling limit (``exact''). Inset: Average Pearson correlation $r_h$ between the true fields and the fields inferred from RMF, same legend.}
\label{fig_sampling}
\end{figure}

\begin{figure}[t!]
\center\includegraphics[width=8cm]{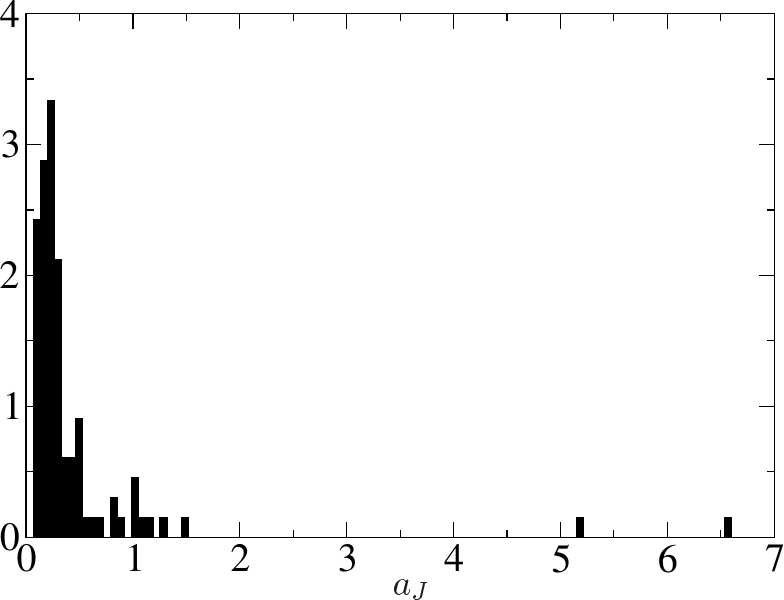}
\caption{Probability distribution of $a_J$  obtained from a hundred sampling of the disorder at $\s_J=0.7$.}
\label{fig_histo_aJ}
\end{figure}

In order to test our approach, we analyze the standard SK model, which consists of $N$ spins interacting with 
random gaussian couplings $J_{ij}$ of zero mean and standard deviation $\s_J=J/\sqrt{N}$, with $N$$=$$10$ spins for $\s_J$$\in$$[0.1,0.9]$ in presence of a random gaussian magnetic field $h_i$, of zero mean and  
standard deviation $\s_h$$=$$0.3$. As the high temperature expansion becomes exact as $N$$\to$$\infty$, a small number of spins is actually 
an interesting test case. We have performed Monte-Carlo simulations for $100$ realizations of the disorders per $\s_J$, and generated various sets for $M$ ranging from $10^2$ to $10^4$ spin configurations, as well as exact computations of the correlation functions, still doable for $N$$=$$10$.  Note that, in the finite sampling cases, our 
simulations were not necessarily thermalized in the strongly correlated regime (i.e. for $\s_J \gtrsim 0.3$), implying possibly strongly biased 
evaluation of the averages, meaning that $\b{f}^M$ and $\bc^M$ can be (and in some cases were) quite different from their exact values for a given realization of the disorder, even for large $M$. These frequencies and correlations were used as input in Eq.~\eqref{result} to obtain the inferred RMF fields and couplings $\bh^*$ and $\bJ^*$. Fig.~\ref{fig_sampling} shows that the quality of the inference is insensitive to sampling if $M$ is large enough, and there is almost no difference between $M=10^3$ and perfect sampling, although $\bc^M$ might be singular, especially at large $\s_J$ (see discussion below). We will thus concentrate on low sampling, which is the most interesting case for biophysics applications. We also find that the couplings are typically of the correct order of magnitude, though smaller than the true ones. That is, if $a_J$ is the slope of the linear regression of $\bJ^*$  against the true $\bJ$, we typically have $a_J\lesssim 1$. We exemplify this in Fig.~\ref{fig_histo_aJ}, which shows the  probability distribution of $a_J$ obtained from a hundred realizations of the disorder at $\s_J=0.7$. We observe that the distribution is picked around $0.25$, but has a long tail. The study of the distribution of $a_J$, for RMF and other inference methods, is an interesting one that we leave for future work.

We now discuss our results for the inference compared to other analytical methods, before showing that the RMF inference is good enough to generate new data, which are similar to the original data.

\subsection{Comparison to other methods}

To assess the validity of our approach, we have also computed the inferred fields and couplings coming from other inference methods, such as the first and second order of the high-
temperature expansion (NMF and TAP), the Bethe approximation (BA), as well as the resummed small correlation expansion of SM \cite{SM09}. The 
corresponding expressions for the coupling can be found for example in \cite{Ricci-Tersenghi2012}. (The expression of fields for the resummed 
small correlation expansion has not been published in the literature, and we have therefore not inferred those.) Note that TAP and BA are based 
on the linear response of the Gibbs free energy, and that for NMF, we have used the ``ring'' results of Sec.~\ref{sec_exp}, which we 
have shown to be equivalent to the so-called NMF (plus linear response) with diagonal matching. For simplicity on the following when we refer to 
NMF we mean this procedure of using the lowest order in the high-temperature expansion plus the linear response method to estimate the 
couplings, plus the diagonal matching method to correct the fields.
Furthermore, all of these approaches involve the inversion of the correlation matrix $\bc$. However,  in the strongly correlated regime, the data are 
very polarized and many  values of pairs of spins are never observed, leading to either rank-deficient, or nearly singular, $\bc^{\rm M}$ matrices, 
due to imperfect sampling.
One way to cure this problem is to use a pseudocount $\a$$=$$2/M$ \cite{BCLM14}, which is the solution shown here. In the case of NMF, we have also 
used two different regularizations (with parameter $\eta$$=$$1$), a $L_2$ regularization \cite{CM12,BCLM14} as well as the somewhat 
different 
regularization of \cite{Andreatta13}. These have not changed the results qualitatively, and are thus not shown. It was argued in \cite{BCLM14} that
large values of pseudo counts or $L_2$ regularizations should be used to compensate for deficiencies of the mean-field approximation, so that we 
also have used a constant regularization $\eta$$=$$0.2 \times M$ for all $M$, without any notable change.
Finally we have also tested the so-called Gaussian model \cite{gaussian} 
with regularization, and found results very similar to those of the regularized NMF+linear response+diagonal matching method, so that we show 
only the latter for simplicity.

In the last few years, it has been understood that in presence of a magnetic field, 
standard inference approaches, such as the high-temperature expansion and Bethe approximation (BA), do not converge (i.e. give complex 
valued fields and couplings) \cite{Ricci-Tersenghi2012}, and we have verified that this is indeed the case here, even in the presence of 
regularization. Indeed, it has been shown that even in the case of three spins in a field, where the calculation can be done by hand and thus with a 
perfect ``sampling'', TAP and BA inference can be ill-defined.
In particular, we have found that both TAP and BA inferences give meaningless results for the value of $\s_h$ used in the simulations, and we will 
therefore not show comparisons with these methods. 
 
 \begin{figure}[t!]
\center\includegraphics[width=8cm]{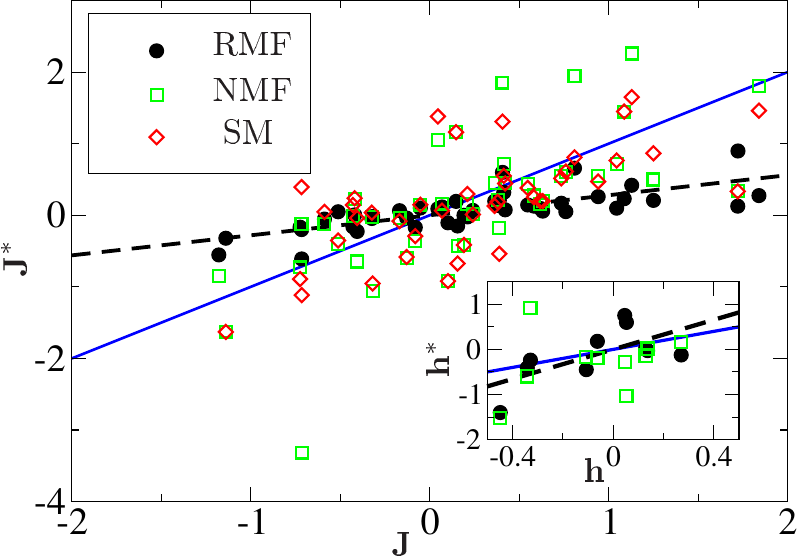}
\caption{Scatter plot of inferred (y-axis) versus true couplings (x-axis), for one typical realization of $\bJ$ and $\b{h}$ at $\s_J$$=$$0.7$, $M$$=$$100$. Black circles: RMF results; green empty squares: NMF; empty diamonds: SM.  Inset: inferred (y-axis) versus true fields (x-axis), same legend (no fields for SM). Dashed lines are linear regressions of the RMF results with slope $0.28$ ($1.65$) for the couplings (fields). Blue full lines have slope $1$. Some NMF and SM points are out of the graph range.}
\label{fig_scatter_SK}
\end{figure}
 
 On the other hand, NMF and SM with pseudocount always allow us to infer couplings (and fields for NMF), although the inference is much less reliable as the temperature decreases, especially at low sampling $M$. On the other hand, our RMF approximation gives limited errors even for  imperfect sampling, see an illustration of this in
Fig.~\ref{fig_scatter_SK}. Note also that we did \emph{not} use any regularization nor pseudocount to perform the RMF inference. We could always converge the matrix $\bD$, that allows us to inverse  $\bD+ \bc^{\rm M}$, even when $\b{c}^{\rm M}$ is rank deficient.

\begin{figure}[t]
\center\includegraphics[width=8cm]{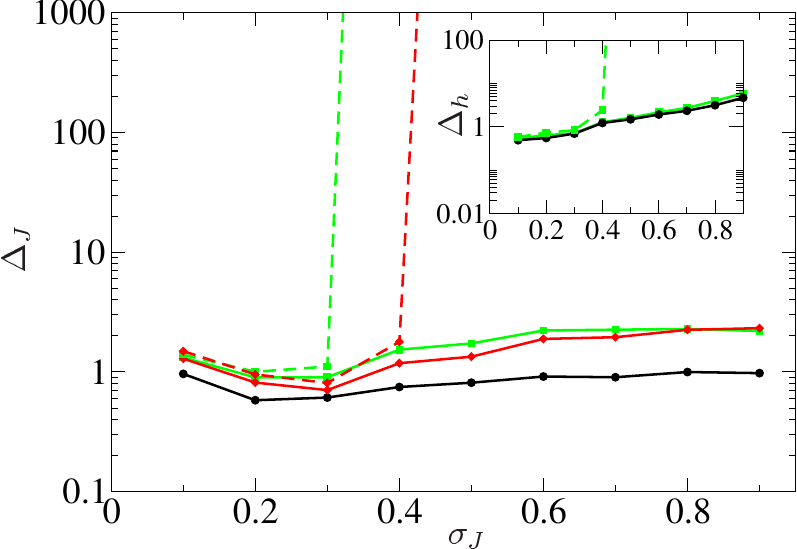}
\caption{Relative error for the fields $\Delta_J$ and couplings $\D_h$ for RMF (black circles), regularized NMF+linear response+diagonal matching (green squares), and regularized SM (red diamonds), averaged over for a hundred realizations of $\bJ$ and $\b{h}$ with $M$$=$$100$. Dashed lines correspond to NMF and SM inference without pseudocount (no regularization or pseudocount is added to RMF). Inset: inferred fields, same legend (SM fields not shown).}
\label{fig_DJDh}
\end{figure}

To quantify the quality of the inference of RMF, NMF and SM, we study two quantities, the relative error
\beq
\Delta_h = \sqrt{\frac{\sum_i (h_i^*-h_i)^2}{\sum_i h_i^2}} \ ,
\eeq 
 as well as Pearson correlations defined by
\beq
\begin{split}
r_h &= \frac{\sum_i\left(h_i-\bar h\right)\left(h_i^*-\bar{h^*}\right)}{\sqrt{\sum_{i}\left(h_i-\bar h\right)^2 \sum_j \left(h_j^*-\bar{h^*}\right)^2}} \ , 
\quad \bar h =\frac{1}{N}\sum_i h_i\ ,
\end{split}
\label{def_pearson}
\eeq
with similar definition for the couplings. 
However, as shown in Fig.~\ref{fig_DJDh}, we find that $\Delta_J$ for RMF typically saturates to one, as the couplings tend to 
be of a   amplitude than the real ones (since $\Delta_J=1$ if $\bJ^*=0$). On the other hand, the the error for the NMF and SM couplings are 
typically large compared to one in the low-temperature regime, for $\s_J\geq 0.3$.  Fig.~\ref{fig_DJDh} also shows the NMF and  SM inference 
without regularization (see dashed lines). In that case, if $\bc^M$ was not invertible, we did not take the corresponding realization into account. 
One clearly sees that this unregularized inference is completely meaningless in the low-temperature regime, and that the use of pseudocount 
improves strongly the results.

\begin{figure}[t!]
\center\includegraphics[width=7cm]{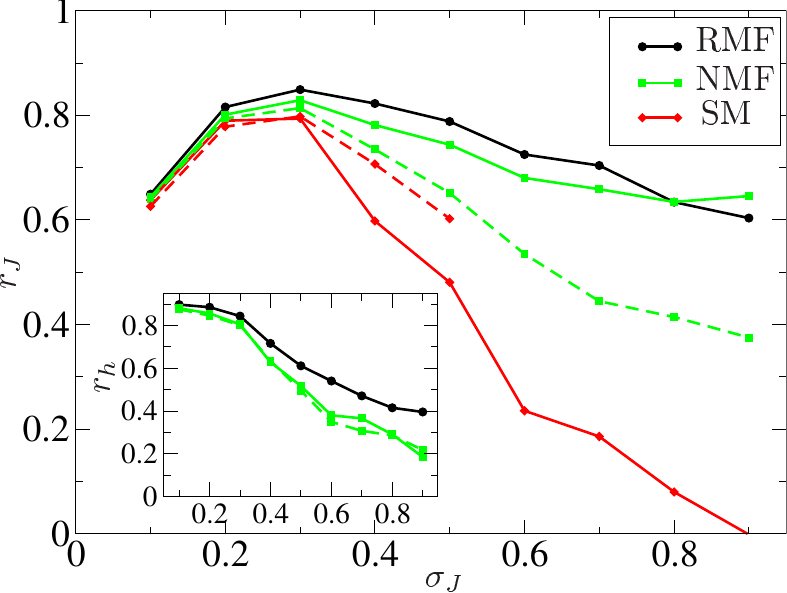}
\caption{Average Pearson correlation $r_J$ between the true couplings and the couplings inferred from RMF (black circles), regularized NMF+linear response+diagonal matching (green squared) and regularized SM (red diamonds). Inset: Pearson correlations $r_h$ between the true fields and the fields inferred from 
RMF and regularized NMF+linear response+diagonal matching. 
The Pearson correlations are averaged over a hundred realizations of $\bJ$ and $\b{h}$.
}
\label{fig_corr_couplings}
\end{figure}

Another way of quantifying the success of the inference is to study the Pearson correlation, that quantifies the correlation between the real and 
inferred couplings, irrespective of the amplitude of the couplings and thus of the error. In particular, one can have a very large error (because all 
inferred couplings are such that $|J^*_{ij}|\gg|J_{ij}|$), but a very good Pearson $r_J\simeq 1$ as the inferred couplings have the correct order or 
ratio between each other. In fact, this is exactly what we observe in Fig.~\ref{fig_corr_couplings}, which shows that for all three methods, the 
Pearson correlations are rather good for all $\s_J$ (although SM seems to break down for $\s_J\gtrsim0.5$).
Note that  Pearson correlations imply that the corresponding interaction graph, as well as the biases distribution, is correct, even if the magnitude 
of the couplings is not well estimated. The observation that NMF (with pseudocount or regularization) gives rather good  Pearson correlations even 
in the low-temperature phase might explain why this method (and its generalization) has been successful to infer the interaction graph in real data.

\begin{figure}[!t]
\center\includegraphics[width=7cm]{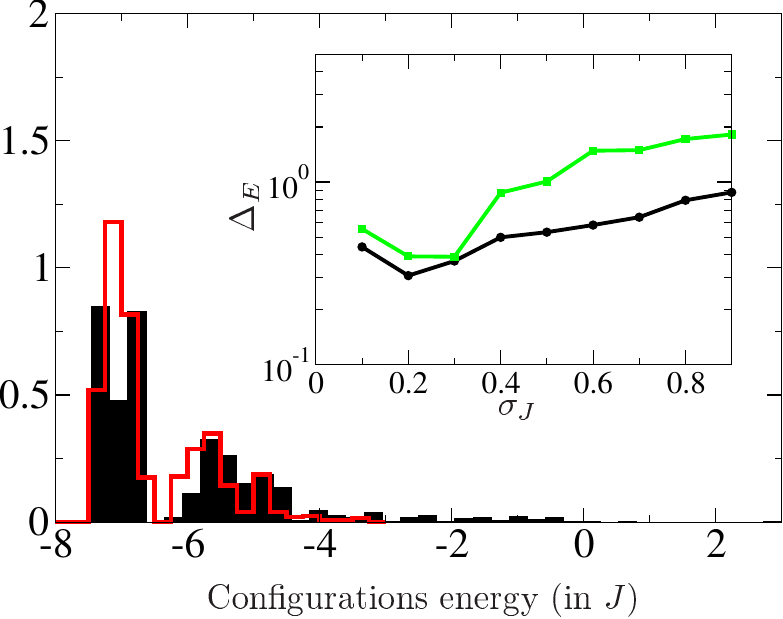}
\caption{Distribution function of the energy $E$($\bJ$,$\b{h}$) of 1000 spin configurations sampled from the true distribution ($\bJ$, $\b{h}$) for the same disorder realization than Fig.~\ref{fig_scatter_SK}, $\{\mathcal{C}\}$ (red histogram), RMF inference  ($\bJ^*$, $\b{h}^*$) with M=100,  $\{\mathcal{C}_{\rm RMF}\}$ (black). 
Inset : Error $\Delta_E$ of the energy obtained from the true and inferred couplings and fields, averaged over the realizations of disorder, see text. Same legend as Fig.~\ref{fig_DJDh}. }
\label{fig_energies}
\end{figure}

However, one has to keep in mind that a good interaction graphs (and biases distributions) are not sufficient to be able to generate new data, typical of the real probability distribution. Indeed, the inferred probability of a given spin configuration $\b{\s}$ is given (up to a constant) by $\exp\left(-E\left(\b{\s};\bJ^*,\bh^*\right)\right)$, where the definition of the energy
\beq
E(\b{\s};\bJ^*,\bh^*) = - \sum_{i} h^*_{i} \s_{i} - \sum_{i<j} J^*_{ij} \s_{i} \s_{j} \  ,
\eeq 
is such that the most probable configurations have the smallest energy. Since the ratio of the probability of two configurations is governed by the difference of the energy,  an inference which has a good interaction graph but a wrong order of magnitude in the fields and couplings will not be able to generate typical configurations (generically, only a few configurations will have a small energy compared to all the other). This point can be exemplified as follow. We have generated a thousand new configurations $\{\mathcal{C}\}$ sampled from the original model $(\bh,\bJ)$, not used for the inference of $(\b{h^*},\bJ^*)$, and computed the energy of each of these configurations $\b{\s}$ with the true couplings and fields $E(\b{\s};\bJ,\bh)$ and the inferred couplings and fields $E(\b{\s};\bJ^*,\bh^*)$ for both RMF and NMF (not SM, since we do not have an expression for the fields). The inset of Fig.~\ref{fig_energies} shows the relative error of the energy,
\beq
\Delta_E=\sqrt{\frac{\sum_{\{\mathcal{C}\}}\left(E\left(\b{\s};\bJ^*,\bh^*\right)-E\left(\b{\s};\bJ,\bh\right)\right)^2}{\sum_{\{\mathcal{C}\}}\left(E\left(\b{\s};\bJ,\bh\right)\right)^2}}\ ,
\eeq
which shows the superiority of RMF over NMF. Thus NMF will be unable to generate meaningful new data, whereas RMF should. Once more, $\Delta_E^{\rm RMF}$ saturates to one because the inferred energy of a given configuration is typically small than that of the true energy (because the couplings are typically smaller), but as we will show now, this still allows us to generate new data that are typical of the original probability distribution.

\subsection{Data generation from RMF inference}

\begin{figure}[t]
\includegraphics[width=8cm]{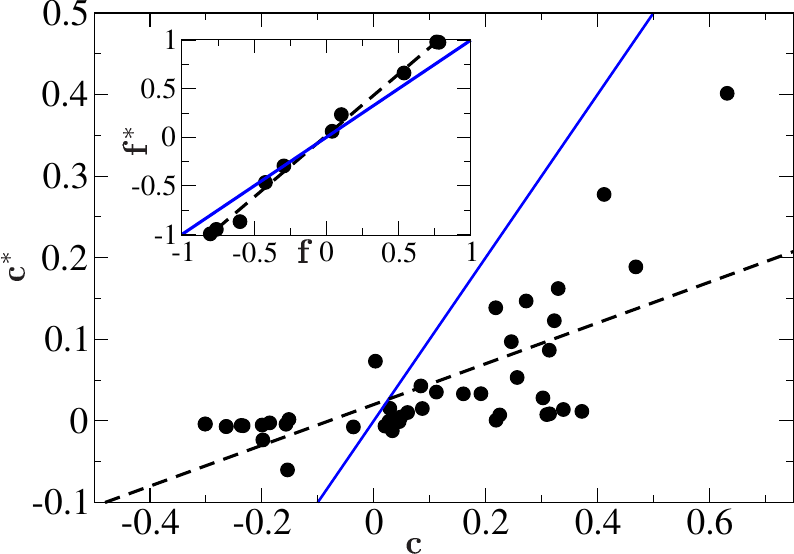}
\caption{Scatter plot of $\b{c^*}$ computed form RMF  inference and $\b{c}$ computed from the original model, for one realization of the disorder (same as in Fig.~\ref{fig_scatter_SK}) for $\s_J=0.7$. Full blue line has slope $1$. Black dashed line is the best linear regression, slope $\simeq 0.25$. Inset: scatter plot for the magnetizations. Slope of the black dashed line $\simeq 1.25$.
}
\label{fig_scatter_cm}
\end{figure}

One of the main interest  of the inference is the ability to generate new configurations, which have high probabilities (i.e. low energy) in the real 
model \cite{SLRLGR05}. Using the RMF couplings ($\bJ^*$, $\b{h}^*$), we have generated a thousand configurations $\{\mathcal{C}_{\rm RMF}\}$ 
via Monte-Carlo sampling. To test whether these configurations would have a high probability, one can compare the energy of each configuration 
of $\{\mathcal{C}_{\rm RMF}\}$ in the original model ($\bJ$, $\b{h}$) to the energy of typical configurations drawn from the real model (see \cite{hugomonasson} for a similar procedure for biological data). We  find that 
the configurations $\{\mathcal{C}_{\rm RMF}\}$ generically have low energy in the real model, i.e., they are configurations that are typical of the original 
model. Fig.~\ref{fig_energies} shows an example of these energy distributions, for the couplings inferred from the same realization of ($\bJ$, $\b{h}$) than 
in Fig.~\ref{fig_scatter_SK}.

\begin{figure}[t]
\includegraphics[width=8cm]{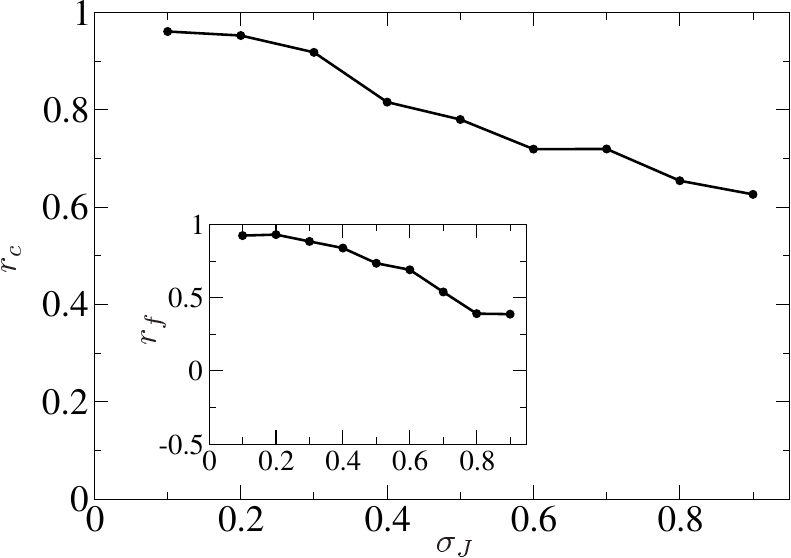}
\caption{Pearson correlations $r_c$ between $\b{c}$ computed form RMF  inference (with $M=100$) and $\b{c}$ computed from the original model, average over a hundred realizations of $\bJ$ and $\b{h}$. The inset shows the average Pearson correlation for the frequencies $r_f$. }
\label{fig_corr_cf}
\end{figure}

Another way to  judge whether the inferred couplings and fields from the RMF approximation are meaningful is to compare the frequencies and correlations obtained using the true and inferred couplings and fields.
To that purpose, we have drawn 1000 configurations via Monte-Carlo sampling, using the original couplings and fields ($\{\mathcal{C}\}$) and the RMF inferred ones ($\{\mathcal{C}_{\rm RMF}\}$) (inference done
with $M$$=$$100$). We have then obtained from these the corresponding frequencies $\b{f}$ and correlations $\b{c}$ (we will denote by ($\b{f}^*$,$
\b{c}^*$) those computed using the inferred couplings ($\b{h}^*$,$\b{J}^*$)). 
We show in Fig.~\ref{fig_scatter_cm} a scatter plot of the ($\b{f}^*$,$\b{c}^*$) vs ($\b{f}$,$\b{c}$) for the same realization of the disorder than in 
Fig.~\ref{fig_scatter_SK}, which shows that the RMF magnetization are really good. Concerning the correlations, we see that RMF gives relatively smaller correlations than the true 
one, which might be understood by the fact that the couplings tend to also be too small. 
We have also computed the corresponding Pearson correlations between ($\b{f}$, $\b{c}$) obtained from $\{\mathcal{C}\}$, and  ($\b{f}^*$, $\b{c}^*
$)  obtained from $\{\mathcal{C}_{\rm RMF}\}$, see Fig.~\ref{fig_corr_cf}. We see that the correlation is rather good even in the low-temperature phase, and thus the RMF can be used to 
generate configurations that indeed reproduce the properties of the real data.

\section{The case of Potts variables \label{sec_potts}}

In this section we generalize our calculation to Potts variables, which is of great interest for biophysical applications. 
We start by considering a Potts model with $q$ possible states for each 
unit, i.e. $\s_{ia}$$=$$\delta_{a \,a_i}$, where $a_i$$=$$1\cdots q$ represents the $q$ possible states on site $i$. 
Switching from Potts to Ising only requires to change the expression for the independent model entropy, and introducing additional
summations over Potts indices. If needed, the generalization to different number of states per site (i.e. $q$$\to$$q_i$) is also straightforward
and not shown here. \\

The temperature dependent partition function reads now
\beq
Z[\bh,\bJ]=\sum_{\b{\s}} \exp \left( \sum_{i,a} h_{ia} \s_{ia} +   \sum_{i<j,a,b}\!\! J_{ia,jb} \s_{ia} \s_{jb} \right) \ .
\eeq
Note that there is  an ambiguity in the way of defining fields and couplings and one has to fix a gauge to remove it.
A simple way to see this issue is to consider the one- and two-point functions
\beq
f_{ia} = \la \s_{ia} \ra\;, \text{ and } \quad p_{ia,jb} = \la \s_{ia} \s_{jb} \ra \ ,
\eeq
which have to satisfy a set of simple constraints,
\beq 
\sum_{a=1}^q f_{ia} = 1  \ , \quad \text{ and } \quad \left\{ \begin{array}{ll}
&\displaystyle \sum_{a=1}^{q} p_{ia,jb} = f_{jb}\ , \\
&\displaystyle \sum_{b=1}^{q} p_{ia,jb} = f_{ia}\ .
\end{array} \right.
\eeq
When trying to infer the fields and couplings we will find that we have too many variables with respect to the set of equations that fix their values. A simple way to fix this is to choose fields and couplings such that
\beq \begin{split}
& h_{iq} = 0 \quad \forall ~ i \ ,\\
& J_{ia,jq} = J_{iq,jb} = J_{iq,jq} = 0 \quad \forall ~ i<j,a,b\ .
\end{split}\eeq
Note that $ J_{ia,ib}=0 \quad  \forall ~ i,a,b$. Other choices are possible, and the calculation can easily be repeated with different gauge, only impacting the 
independent model Gibbs free energy. In the following, all summations over Potts indices will thus run from $1$ to $q-1$ unless specified 
otherwise. Finally we will often gather the $N(N-1)/2 \times (q-1)$ parameters $J_{ia,jb}$ in a matrix $\b{J}$, in which the $i=j$ elements are $0$, the $i<j,a,b$ 
elements are $J_{ia,jb}$, and the $i>j,a, b$ elements are $J_{jb,ia}$.

The entropy of the independent model reads now
\beq\begin{split}
\SS_{\rm IM}[\b{f}] = & ~  - \sum_{i} \sum_{a} f_{ia} \ln f_{ia} \\
& - \sum_{i} \left( 1 - \sum_{a} f_{ia} \right) \ln \left( 1 - \sum_{a} f_{ia} \right)\ ,
\end{split}\eeq
and we will need, as in the Ising case, its matrix of second derivatives
\beq
\left(\bL^{-1}\right)_{ia,jb} = - \frac{\d^2 \SS_{\rm IM}}{\d f_{ia} \d f_{jb}} = \d_{ij} \frac{1}{f_{ia}} \left( \d_{ab} + \frac 1{1- \sum_b f_{ib}} \right) \  ,
\label{invL_potts}
\eeq
which is the inverse of the self-correlation matrix $\b{L}$ given by
\beq
L_{ia,jb} = \d_{ij} f_{ia} \left( \d_{ab} - f_{ia} f_{ib} \right) \  .
\eeq
We also need the higher-order derivatives of $\SS_{\rm IM}$ which read
\beq
\g^{(n)}_{i_1 a_1,\cdots,i_n a_n} = - \frac{\d^n \SS_{\rm IM}[\b{f}]}{\d f_{i_1 a_1} \cdots \d f_{i_n a_n}} \   .
\label{vertices_potts}
\eeq
All steps of the derivation of the RMF approximation are now the same, and we obtain
\beq\begin{split}
\GG_{\e}[\b{f},\bJ]  = & ~  \SS_{\rm IM}[\b{f}] +  \sum_{i<j,a,b} J_{ia,jb} f_{ia} f_{jb} \\
& - \frac{\e}{2} \Tr \left[ \ln \left( \bL^{-1} -    \bJ \right) - \ln \bL^{-1} \right] + \OO(\e^2)\ ,
\end{split} \eeq
from which the equation for the RMF couplings is obtained
\beq
 J^*_{ia,jb} = - \left( \b{c} + \b{D} \right)^{-1}_{ia,jb} \quad \forall ~ i<j,a,b \ .
\label{eq_JPotts}
\eeq
The $\b{D}$ matrix is now defined by the $N \times (q-1)^2$ coupled equations:
\beq
\left( \b{c} + \b{D} \right)^{-1}_{ia,ib} = L^{-1}_{ia,ib} \quad \forall ~  i,a,b\ ,
\label{eq_DPotts}
\eeq
and the expression for the RMF fields is now
\beq \begin{split}
h^*_{ia} = & ~  \ln \left( \frac{f_{ia}}{1-\sum_b f_{ib}} \right) - \sum_{j (\ne i),b}  J^*_{ia,jb} f_{jb} \\
& + \frac{\e}{2} \sum_{b,c} D_{ib,ic} \g^{(3)}_{ia,ib,ic} \ .
\end{split}\eeq
Finally the expression for the entropy for Potts variables is
\beq\begin{split}
\SS_{\e}[\b{f},\b{p}] = & ~ \SS_{\rm IM}[\b{f}] - \frac \e2 \Tr \left( \b{D} \bL^{-1} \right) \\
& + \frac{\e}{2} \Tr \left[ \ln \left( \b{c} + \b{D} \right) - \ln \bL \right] + \OO(\e^2) \ .
\end{split}
\label{RMF_entropy_potts}
\eeq \\

\section{Expansion to second order in $\e$ \label{sec_2ndorder}}

Before plunging into the next order calculation, a remark is in order. Although we want to illustrate, by pushing to the next order, that our
approximation scheme is systematic, we might expect little improvement for realistic data. Indeed, higher order terms 
 involve a large number of summations over Potts and site indices, leading to a greater numerical 
sensitivity to sampling noise and to a larger complexity of the calculation. 
On the other hand the entropy functional will be better approximated using this second order approximation, 
but this improvement will most probably be impaired by sampling noise. The question of quantifying the interplay between these two effects
is of interest, but we leave it for future work. 

\begin{figure}[t!]
\includegraphics[width=8cm]{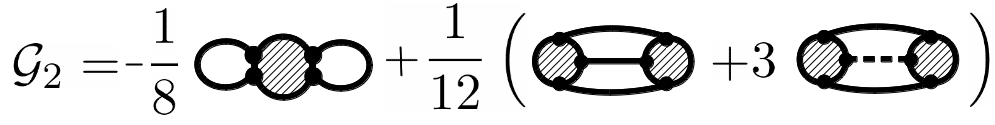}
\caption{Diagrammatic representation of the second order correction to the Gibbs potential. Grey blobs with $n$ dots represent $\g^{(n)}$ vertices. Black dots represent summations over the indices, full lines represent $\bG(\beta)$ and dotted lines represent $\bL$.}
\label{fig_diag2}
\end{figure}

In order to go to the next order, we define the propagator $\b{G}(\be)$ as
\beq
G_{ia,jb}(\be) = \left( \bL^{-1} - \be \bJ \right)^{-1}_{ia,jb} - L_{ia,jb}
\eeq
The expansion in powers of $\e$ of the correlation function is shown in Eq.~\eqref{exp_c} 
and using it in the equivalent of Eq.~\eqref{Wetterich} for Potts variables leads to the equation for $\GG^{(2)}$,
\beq
 \GG^{(2)} =\frac 12 \int_0^1 d\be  \sum_{ia,jb} J_{ia,jb} \left[ \bch^{(0)} \frac{\d^2 \GG^{(1)}}{\d \b{f} \d \b{f}}
\bch^{(0)} \right]_{ia,jb} \ .
\label{eq_GG2}
\eeq
We need to compute the derivative of $\GG^{(1)} $. For compactness, we gather pairs of indices like $i,a$ or $j,b$ in greek letters $\a$,$\g$. 
We find
\beq\begin{split}
& \frac{\d^2 \GG^{(1)}[\b{f},\be \bJ]}{\d f_{\a} \d f_{\g}} = - \frac 12 \sum_{\m,\n} \g^{(4)}_{\a, \g, \m, \n} G_{\m\n}(\be) \\
& + \frac 12 \sum_{\m,\n,\d,\o} \g^{(3)}_{\a,\m,\n} \left[ G_{\m\d}(\be) G_{\n\o}(\be) + 2 L_{\m\d} G_{\n\o}(\be) \right] \g^{(3)}_{\d,\o,\g}\ ,
\end{split}\eeq
which shows that the dependance on $\be$ in this expression is only through $\b{G}(\be)$. We remark that for any
functional $F[\b{G}(\be)]$ we have
\beq\begin{split}
\partial_\be F[\b{G}(\be)] & = \sum_{\a,\g} \frac{\d F}{\d G_{\a\g}} \partial_\be G_{\a\g}(\be) \\
& = \sum_{\a,\g,\mu,\nu} J_{\a\g} ~ \chi^{(0)}_{\a\mu} ~ \frac{\d F}{\d G_{\mu\nu}} ~ \chi^{(0)}_{\nu\g} \ .
\end{split}\eeq
Coming back to Eq.~\eqref{eq_GG2}, we see that we only have to integrate the second derivative of $\GG^{(1)} $ 
with respect to $\b{G}(\be)$ in order to put the r.h.s. in the form of a total derivative w.r.t. $\be$. Defining
the functional
\beq \begin{split}
F^{(2)}[\b{G}] = & ~ - \frac 14 \sum_{\a,\g,\mu,\nu}  G_{\a\g}\g^{(4)}_{\a,\g,\mu,\nu} G_{\mu\nu} \\
&+ \frac 16 \sum_{\a,\g,\d,\l,\mu,\nu} \g^{(3)}_{\a,\g,\d} G_{\a\mu} G_{\g\nu} G_{\d\l}  \g^{(3)}_{\mu,\nu,\l}\\
& + \frac 12  \sum_{\a,\g,\d,\l,\mu,\nu} \g^{(3)}_{\a,\g,\d} G_{\a\mu} G_{\g\nu} L_{\d\l}  \g^{(3)}_{\mu,\nu,\l}\ ,
\end{split}\eeq
we have 
\beq
\frac{\d F^{(2)}[\b{G}]}{\d G_{ia,jb}} = \frac{\d^2 \GG^{(1)} }{\d f_{ia} \d f_{jb}}\ ,
\eeq
so that the equation for $\GG^{(2)} $ is now easily integrated w.r.t. $\be$ to give
\beq
\GG^{(2)} [\b{f},\bJ] = \frac 12 F^{(2)}[\b{G}(1)] \ .
\eeq
This expression can be represented  diagrammatically as shown in Fig.~\ref{fig_diag2}. 
In the case of Ising variables, the final expressions largely simplify due to the locality of the vertices $\g^{(n)}$ and the overall absence of Potts indices, and
in that case we have been able to push the expansion to the third order in $\e$. However the number of diagrams involved rapidly increases beyond that point.
Another issue, already discussed above, is that high order terms in the expansion involve several matrix products of correlation functions, which might render the numerical scheme very sensitive
to sampling noise, which is the reason why we have  only shown in this paper numerical tests of the lowest order. \\

We can now deduce the second order contribution to the optimal couplings.
We set
\beq
J^*_{ia,jb} = J^{(1)}_{ia,jb} + \e J^{(2)}_{ia,jb} + \OO(\e^2)\ ,
\label{eq_Jeps2}
\eeq
and insert this expansion in the optimization equation obtained through the differentiation of $\GG^{(2)}$ w.r.t. $\bJ$:
\beq\begin{split}
c_{ia,jb} = & ~ \left( \bL^{-1} -  \bJ^* \right)^{-1}_{ia,jb} 
+ \frac{\e}{2} \frac{\d F^{(2)}}{\d   J_{ia,jb}} + \OO(\e^2)\ .
\end{split}
\label{eq_ceps2}
\eeq
Inverting Eqs.~\eqref{eq_Jeps2} and \eqref{eq_ceps2} order by order in $\e$, we find
\beq \begin{split}
& \b J^{(2)}_{ia,jb} =  \frac 12 \sum_{\m,\n} \g^{(4)}_{ia,jb,\m,\n} \left( \b{c} + \b{D} - \bL \right)_{\m\n} \\
& - \frac 12 \sum_{\m,\n,\d,\o} \g^{(3)}_{ia,\m,\n} \left[ \left( \b{c} + \b{D} \right)_{\m\d}\left( \b{c} + \b{D} \right)_{\n\o}-\bL_{\m\d} \bL_{\n\o} \right] \g^{(3)}_{\d,\o,jb} \\
& - \left[ \left( \b{c} + \b{D} \right)^{-1} \b{D}^{(2)} \left( \b{c} + \b{D} \right)^{-1} \right]_{ia,jb} \ ,
\end{split}\eeq
where $\b{D}^{(2)}$ is a block diagonal matrix (playing the role of $\bD$ to the next order in $\e$),
\beq
\begin{split}
& D^{(2)}_{ia,ib} = - \sum_{j,c,d} A^{-1}_{iab,jcd} H^{(2)}_{jc,jd} \ ,\\
& A_{iab,jcd} = \left( \b{c} + \b{D} \right)^{-1}_{ia,jc} \left( \b{c} + \b{D} \right)^{-1}_{ib,jd}  \ ,\\
& H^{(2)}_{ia,ib} = - \frac 12 \sum_{c,d} \g^{(4)}_{ia,ib,ic,id} D_{ic,id} \\
& + \frac 12 \sum_{c,d,e,f} \g^{(3)}_{ia,ic,id} \left[ D_{ic,id} D_{ie,if} + 2 L_{ic,id} D_{ie,if} \right] \g^{(3)}_{ib,ie,if}\ .
\end{split}
\eeq
Using this result, one can deduce the expression of the inferred fields and the entropy in the spirit of what was done at the lowest order.

\section{Conclusion \label{sec_concl}}

We have introduced the resummed mean-field approximation  for the inference problem in the context of Ising and Potts variables, which is based 
on an exact equation for the Gibbs free energy. At the lowest non-trivial order, we obtained a simple analytical expression for the couplings and 
the fields as functions of the correlations and frequencies. The main difference compared to other approaches is that it does not rely on inversions of the correlation matrix $\bc$, thanks to the 
 matrix $\b{D}$, which is fixed by the dataset itself, implying that RMF works even when $\bc$ is rank deficient, as often happens in real data. The RMF approximation we have obtained corresponds to a resummation of an 
infinite number of terms of the small correlation expansion, and we have shown that it can be pursued in a principled and systematic way.

We have tested the method on the SK model and shown that it works well even in the strongly coupled regime, 
in particular in presence of a magnetic field, where 
other methods break down. The inferred couplings and fields are well correlated with the real ones, and of the correct order of magnitude, 
although the couplings tend to be 
smaller than expected. A striking result is that we do not need to include a pseudocount or a regularization even for small sampling, or when the 
correlation matrix $\bc^M$ is not invertible, which is a clear improvement upon other mean-field methods. 
In particular, the matrix $\b{D}$ depends only of the data, and prevent the need of optimizing over additional parameters such as a 
pseudocount. This feature is crucial for practical applications, and we expect this, together with the fact that the inference is reliable even 
for large couplings, to pave the way for systematic applications to very large datasets, and/or datasets with units assuming a large 
number of possible states. We have also demonstrated that the inference performed by our method is consistent with the original model at the 
level of the probabilities of single configurations, a feature that could have important implications in bio-informatics \cite{RLMYR05,SLRLGR05}. 
Our method, while being analytic, and hence very fast, outperforms the competing analytical schemes, even when they are regularized.
If necessary our method can also be regularized by adding an $L_1$ or $L_2$ prior on $\b{J}$. For the $L_2$ case, 
Eq.~\eqref{GG_reg} can, for example, be modified to take into account the $\e$ parameter by defining the regularized entropy
\beq
\SS_{\e,{\rm reg}}[\b{f},\b{p}] = \underset{\bJ}{{\rm inf}} \left( \begin{array}{ll}
& \displaystyle \GG_{\e}[\b{f},\bJ] - (1-\e) \sum_{i<j} J_{ij} f_i f_j \\
& \\
& \displaystyle - \e \sum_{ij} J_{ij} p_{ij}
- \frac{\e \eta}{M} \sum_{i<j} J_{ij}^2
\end{array} \right)
\eeq
which leads to the equation for the couplings
\beq
c_{ij} = \left( \bL^{-1} -  \bJ^* \right)^{-1}_{ij} + \frac{\eta}{M} J^*_{ij} \quad \forall ~ i<j \  .
\eeq
If one assumes that $\eta$ is of order $1$, this can be expanded around the $\eta/M=0$ case if $M$ is large enough. 
For the standard Gaussian and NMF methods, 
it was found that the optimal $L_2$ regularization is not of order $1$ but of order $M$, which indeed compensates for the 
deficiencies of these methods. In our case, since the unregularized inference is already well-behaved, we expect that
the optimal regularization will be $\OO(1)$. We leave this issue for future work.

The transition from liquid to spin glass, that comes about (in the direct problem, 
where $\bJ$ is fixed) when the couplings becomes of order one could be thought to be a limit to mean-field inference, due to the apparition
of multiple minima in $\GG$, that prevent correct thermalization of observables like $\s_i \s_j$ \cite{NB12}.
Although the presence of an underlying phase transition (in the thermodynamic limit) could seem to be a hindrance to 
the success of the inference, it has been argued in \cite{CM12}
that this should not be a limitation. The intuitive argument is that since the inverse problem is characterized by the 
inverse susceptibility (i.e. how fields and couplings are affected by a change in magnetizations and correlations), the 
inference should not be hindered 
by a divergence of the susceptibility due to the phase transition, since the inverse susceptibility will stay 
well-behaved. 
Of course, probing deep inside a low temperature phase will lead to data that are very polarized, resulting in bad 
statistical estimation of correlations, but this problem also affects data that have very small correlations, and is not 
tantamount of an underlying thermodynamic phase 
transition. Indeed, numerical methods like the ACE or pseudo-likelihood are not particularly affected by transitions 
towards low-temperature phases \cite{CM11,AE12}. However, the only analytical method prior to our work that focuses on the 
entropy $\SS$ instead of $\GG$ while extending NMF, namely the small correlation expansion of SM, was 
found to also hit the spin glass limit, and was also shown to be extremely sensible to sampling noise \cite{SM09}. 
Our method solves this apparent contradiction and provides an analytical scheme that is unaffected by 
the phase transition, without having to resort to gradient descent to evaluate the couplings, as in the ACE or PLM methods. 

In retrospect, we can understand why resumming closed ring diagrams is important in the finite sampling case 
by introducing sampling noise in the computation.
Since the connected correlation function is the empirical covariance of a vector of $N$ (non independent) variables $\s_1,\ldots,\s_N$,
the finite sampling effects can be taken into account by considering that $\b{\G}$ is sampled from the ensemble of Wishart matrices 
\cite{bouchaud_wishart,dario}. If we consider instead, in a schematic way, that the empirical $\b{\G}^{\rm M}$ matrix is the sum of 
the perfect sampling result $\b{\G}$ and of a matrix of uncorrelated Gaussian elements, with variance $\a$, we see that the closed
loop diagrams that are resummed by our RMF approximation pick $\OO(\a N)$ contributions to the entropy, and should thus not be
neglected in the presence of sampling noise. A more careful analysis of these finite sampling effects will be discussed elsewhere.

\begin{figure}[t!]
\includegraphics[width=8cm]{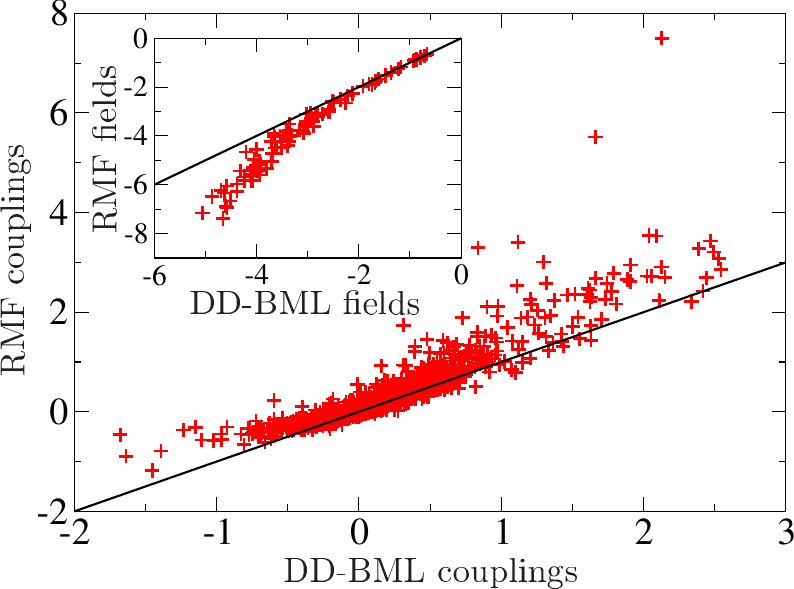}
\caption{Couplings inferred from the RMF approximation, compared to the exact solution to the inference problem, obtained
with the data-driven Boltzmann machine learning algorithm (DD-BML) introduced in \cite{Ferrari15}. The Pearson correlation between the couplings of the two inferences is $0.90$. The dataset is comprised
of $95$ neurons \cite{Mora2015}.  Inset: comparison of the inferred fields, with Pearson correlation $0.97$. Full dark lines have slope $1$.}
\label{fig_neurons}
\end{figure}

An important question for applications to realistic data is that of scalability. 
Although we have tested our method on the SK model with a small number of spins, it can be used for much larger numbers of units.
In order to demonstrate the ability of our method to go beyond toy models, we show in Fig.~\ref{fig_neurons} the inferred couplings obtained with the 
RMF method when analyzing neuronal data of the retina of a rat, taken from Ref.~\cite{Mora2015}. 
The RMF couplings are compared to the couplings obtained through Boltzmann machine learning (computed with the method of 
\cite{Ferrari15}), that provides the exact solution to the inference problem (the magnetizations and correlations are perfectly reproduced
by the model inferred through this method). We see that the agreement between the RMF couplings and the Boltzmann machine learning
couplings is very good. In that case the number of neurons was $N=95$ and the number of configurations $M=485998$. 
The RMF algorithm has converged in a split second on a personal desktop computer. 
This very large 
number of samples ensures that the data-driven Boltzmann machine learning algorithm (DD-BML in the figure) 
of \cite{Ferrari15} solves the inference problem exactly, providing a ground truth for comparison. 
Such large sampling is of course not mandatory for our method to converge.

Finally we also have inferred RMF couplings from a notMNIST dataset comprised of $M=5000$ images from which $N=784$ pixels 
were extracted. The inference also took around a second on a personal computer. No ground truth is known in that case, so we leave the analysis 
of these results for future work, but this demonstrates that the method scales favorably with the number of interacting units in the data.

Our method presented here can be easily generalized to other kind of variables, e.g. continuous variables, along the lines of \cite{Machado2010}, or to 
restricted Boltzmann machines following \cite{GTK15}. The main requirement for the method to be valid is that there exist a well defined Independent Model Gibbs functional 
$\GG_{\rm IM}$. One could also imagine dealing with quantum variables in the context of quantum inference, following the calculation of \cite{Rancon2014}. \\

\acknowledgments 

H.J. was funded by the Agence Nationale de la Recherche Coevstat project (ANR-13-BS04-0012-01) in early stages of this work, 
and acknowledges support from the Laboratoire de Physique Statistique de l'\'Ecole Normale Sup\'erieure de Paris. A.R. was funded by 
ANR ``ArtiQ'' project.
We thank C.K. Fisher and U. Ferrari for stimulating discussions, F. Krzakala and R. Monasson for important suggestions, and S. Deny and 
N. Dupuis for useful comments. We warmly thank O. Marre, U. Ferrari and S. Deny for providing the data shown in Fig.~\ref{fig_neurons}.


\appendix

\section{Alternative derivation of RMF \label{app_saddle}}

The expansion in $\e$  is equivalent to the so-called ``loop expansion'' used in field theory, usually interpreted as a saddle point plus fluctuations expansion of a given functional integral \cite{zinn}.
We give for completeness this alternative derivation of our results, connecting it with other
works \cite{PP95, CMV11}, while helping the reader to get more intuition about the role of the
parameter  $\e$ that is used to organize the expansion around NMF.
Starting from the Ising partition function we perform a Hubbard-Stratonovich transformation
which leads to
\beq\begin{split}
& Z[\bh,\bJ] = \frac 1{\sqrt{\det \b{J}}} \int \DD \b{\phi} ~ e^{- S[\b{\phi};\bh,\bJ]} \ ,\\
& S[\b{\phi};\bh,\bJ] = \frac 12 \sum_{i,j} \phi_i \left( \b{J} \right)^{-1}_{ij} \phi_j
- \sum_i \ln 2 \cosh \left( h_i + \phi_i \right) \ .
\end{split}\eeq
Cases where $\bJ$ is not definite positive can be dealt with by a suitable shift on its diagonal, see for instance \cite{Machado2010}.
Starting from this representation, one could once more perform a small-coupling expansion  as done in \cite{PP95}, and recover the results of Ref.~\cite{GY91}.

Instead, one can also perform a saddle point approximation, plus fluctuations. To do so, one introduces a small parameter $\e$, such that the saddle point becomes exact in the limit $\e\to0$,
\beq \begin{split}
& Z_{\e}[\bh,\bJ] = \frac 1{\sqrt{\det \b{J}}} \int \mathcal{D} \b{\phi} ~ \exp \left( - \frac 1{\e} S[\b{\phi};\bh,\bJ] \right) \ .
\end{split}\eeq
One can then perform the first Legendre transform of $\ln Z_\e$ with respect to the fields $\bh$,
\beq
\GG_\e[\b{f},\bJ] = \underset{\bh}{{\rm inf}}  \left( \e \ln Z_\e[\bh,\bJ] - \sum_{i=1}^N h_i f_i \right)
\eeq
Expanding $\GG_\e$ in powers of $\e$ leads to Eq.~\eqref{eq_GG1loop}, while performing the second Legendre transform with respect to $\bJ$ afterwards leads to the RMF. This gives an interpretation of our expansion in terms of the saddle-point evaluation of a path integral representation of the Ising model.

This type of path integral representation is well-known and
was already used in the direct problem, for example in \cite{PP95} and in the inverse problem for example in \cite{CMV11} albeit with a
particular choice of form for the $\bJ$ matrix.

The approach we have used in the main text has been shown (albeit for a more standard $\phi^4$ theory) to be equivalent to the loop expansion \cite{Litim2002}. However, it  has several advantages. First, it allows one to work directly with the microscopic degrees of freedom, which makes it
very easy to generalize to other kind of variables, while avoiding the presence of the complicated $\ln\cosh(\phi+h)$ potential. Furthermore, the
exact equation directly involves $\GG$, which implies that only one Legendre transform (with respect to $\bJ$) has to be performed, which greatly
simplifies the calculation if one wants to push the expansion to higher order in $\e$.

\section{High-temperature expansion from Wetterich equation \label{app_exp}}

In this appendix, we show how to recover the high-temperature (small $\be$) expansion of the Gibbs free energy $\GG[\b{f},\be\bJ]$ of the Ising model, developed by Plefka to order $\be^2$ \cite{Pl82}, and obtained up to order $\be^4$ by Georges and Yedidia (GY) \cite{GY91}. One can then obtain the small correlation expansion (i.e. the expansion of $\SS[\b{f},\b{p}]$ for small $\b{\tilde{c}}$) at a given order in $\b{\tilde{c}}$ by performing explicitly the Legendre transform of $\GG[\b{f},\be\bJ]$ at the same order in $\be$.

\begin{figure}[t!]
\includegraphics[width=8cm]{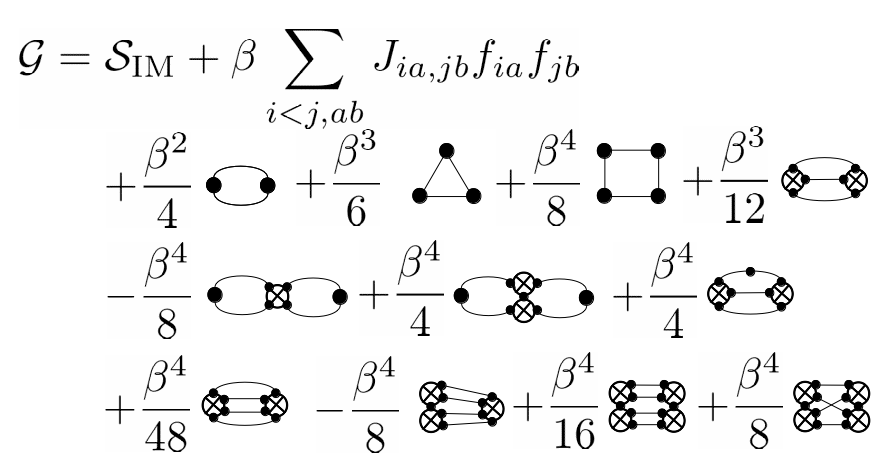}
\caption{High-temperature expansion of $\GG_{\be}$ at order $\be^4$. A black dot represents a summation over one pair of
indices (site index and color index), a black line is a $\bt{J}$ matrix as defined in Eq.~\eqref{rescaled_J_gamma},
and a crossed vertex with $n$ attached dots represents a $V^{(n)}$ vertex
as defined in Eq.~\eqref{rescaled_J_gamma}.}
\label{diagrams_gibbs_potts_O4}
\end{figure}

We start with the Wetterich equation, Eq.~\eqref{Wetterich}, and expand both $\GG[\b{f},\be\bJ]$ and $\bch$ in $\be$,
\beq
\begin{split}
 \GG[\b{f},\be\bJ] &= \SS_{\rm IM} +\be \GG_{1}+\be^2 \GG_{2}+\be^3 \GG_{3}+\be^4 \GG_{4}+\OO(\be^5)\ , \\
 \bch&=\bL+\be\bL\frac{\d^2 \GG_{1}}{\delta\b{f}\delta\b{f}}\bL\\
 &+\be^2\left(\bL\frac{\d^2 \GG_{2}}{\delta\b{f}\delta\b{f}}\bL+\bL\frac{\d^2 \GG_{1}}{\delta\b{f}\delta\b{f}}\bL\frac{\d^2 \GG_{1}}{\delta\b{f}\delta\b{f}}\bL\right)\\
 &+\be^3\bigg(\bL\frac{\d^2 \GG_{3}}{\delta\b{f}\delta\b{f}}\bL+\bL\frac{\d^2 \GG_{2}}{\delta\b{f}\delta\b{f}}\bL\frac{\d^2 \GG_{1}}{\delta\b{f}\delta\b{f}}\bL\\
 &+\bL\frac{\d^2 \GG_{1}}{\delta\b{f}\delta\b{f}}\bL\frac{\d^2 \GG_{2}}{\delta\b{f}\delta\b{f}}\bL+\bL\frac{\d^2 \GG_{1}}{\delta\b{f}\delta\b{f}}\bL\frac{\d^2 \GG_{1}}{\delta\b{f}\delta\b{f}}\bL\frac{\d^2 \GG_{1}}{\delta\b{f}\delta\b{f}}\bL\bigg)\\ &+\OO(\be^4)\ ,
 \end{split}
\eeq
where $L_{ij}=\d_{ij}(1-f_i^2)=\d_{ij}L_i$. Using this expansion and Eq.~\eqref{Wetterich} gives the hierarchy of equations (using the fact that $\bL$ is diagonal and $\bJ$ is zero on the diagonal),
\beq
\begin{split}
 \GG_{1}&=\sum_{i<j} J_{ij} f_i f_j \ , \\
 \GG_{2}&=\frac14{\rm Tr}\left(\bL\bJ\bL\frac{\d^2 \GG_{1}}{\delta\b{f}\delta\b{f}}\right) \ , \\
 \GG_{3}&=\frac16{\rm Tr}\left(\bL\bJ\bL\frac{\d^2 \GG_{2}}{\delta\b{f}\delta\b{f}}+\bL\bJ\bL\frac{\d^2 \GG_{1}}{\delta\b{f}\delta\b{f}}\bL\frac{\d^2 \GG_{1}}{\delta\b{f}\delta\b{f}}\right)  \ , \\
 \GG_{4}&=\frac{1}{8}{\rm Tr}\bigg(\bL\bJ\bL\frac{\d^2 \GG_{3}}{\delta\b{f}\delta\b{f}}+\bL\bJ\bL\frac{\d^2 \GG_{2}}{\delta\b{f}\delta\b{f}}\bL\frac{\d^2 \GG_{1}}{\delta\b{f}\delta\b{f}}\\
 &+\bL\bJ\bL\frac{\d^2 \GG_{1}}{\delta\b{f}\delta\b{f}}\bL\frac{\d^2 \GG_{2}}{\delta\b{f}\delta\b{f}}+\bL\bJ\bL\frac{\d^2 \GG_{1}}{\delta\b{f}\delta\b{f}}\bL\frac{\d^2 \GG_{1}}{\delta\b{f}\delta\b{f}}\bL\frac{\d^2 \GG_{1}}{\delta\b{f}\delta\b{f}}\bigg)\ ,\\
 &\vdots
\end{split}
\eeq
From $ \GG_{1}=\sum_{i<j} J_{ij} f_i f_j$ we find
\beq
\frac{\d^2 \GG_{1}}{\delta\b{f}\delta\b{f}}=\bJ\ ,
\eeq
implying
\beq
\GG_{2}=\frac14\sum_{i\neq j}L_i J^2_{ij}L_j \ .
\eeq
Continuing in the same vein, we obtain
\beq
\begin{split}
\GG_{3}&=\frac13\sum_{i\neq j}m_i L_i J^3_{ij}L_j m_j+\frac16\sum_{i\neq j\neq k}L_i J_{ij}L_j J_{jk} L_k J_{ki} \ , \\
\GG_{4}&=\frac{1}{12}\sum_{i\neq j}(1-3m_i^2) L_i J^4_{ij}L_j (1-3m_j^2)\\ &+\sum_{i\neq j\neq k}m_i L_i J_{ij}^2L_j m_j J_{ik} L_k J_{kj}
 - \frac14 \sum_{i\neq j\neq k}L_j J_{ij}^2L_i^2 J_{ik}^2 L_k \\ &+\frac18\sum_{\substack{i,j,k,l \\ i\neq j, j\neq k \\ k\neq l, l\neq i}}L_iJ_{ij}L_jJ_{jk}L_k J_{kl}L_l J_{li}\ .
\end{split}
\eeq

These results are in perfect agreements with that of GY (up to a global sign from the definition of $\GG$), once rewritten in terms of n-uplets of different spins. Note however that our derivation is quite different from that of GY which is based on the explicit evaluation of high order correlations of the independent model such as $\langle \s_i \s_j \s_k \s_l \cdots\rangle$, up to eight spins for the order $\be^4$, which can be cumbersome to evaluate since one has to take into account if the spins are on the same sites or not. On the other hand, our derivation is straightforward and can be easily pushed to higher order, and is generalizable to Potts variables. We have also checked explicitly that our $\e$ expansion to order $\e^3$ allows us to recover the $\beta$ expansion to order $\beta^4$ exactly.

\begin{figure}[t]
\includegraphics[width=8cm]{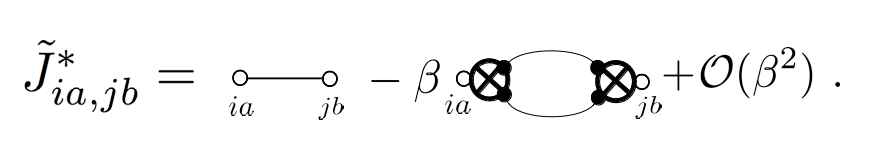}
\caption{Diagrammatic representation of the small correlation expansion of $\bt{J}^*$.
A black dot represents a summation over one pair of
indices (site index and color index), a black line is a $\b{\G}$ matrix as defined in Eq.~\eqref{def_Gamma},
and a crossed vertex with $n$ attached dots represents a $V^{(n)}$ vertex
as defined in Eq.~\eqref{rescaled_J_gamma}.}
\label{fig_diags_Jtilde_potts}
\end{figure}

The case of Potts variables can be treated along the same lines. However the presence of the color indices complicates a lot the summations
and it is best to stick to a diagrammatic representation. To have compact results we define rescaled versions of the
coupling matrix and of the vertices $\g^{(n)}$,
\beq\begin{split}
& \bt{J} = \bL^{1/2} \bJ \bL^{1/2} \  , \\
&V^{(n)}_{i a_1,\ldots,i a_n} = \!\!\! \sum_{j,b_1,\ldots, b_n} \g^{(n)}_{j b_1,\ldots,i b_n} (\bL^{1/2})_{j b_1,i a_1} \!\!\! \cdots (\bL^{1/2})_{j b_n,i a_n} \  .
\label{rescaled_J_gamma}
\end{split}\eeq
We find the high temperature expansion of $\GG $ at order four to be given by Fig.~(\ref{diagrams_gibbs_potts_O4}).

To perform the small correlation expansion using the auxiliary parameter $\be$, one sets the second Legendre transform with a modified 
expression
\beq
\SS[\b{f},\bt{c}] = \underset{\bJ}{{\rm inf}} \left( \GG[\b{f},\be \bJ] - \sum_{i<j} J_{ij} \left[ f_i f_j + \be \t{c}_{ij} \right] \right)
\eeq
The equation setting the optimal couplings, which we show only at second order in $\be$
since we do not need more to obtain the $\OO(\be^4)$ expression for the entropy, is
\beq
\t{J}^*_{ia,jb} = \G_{ia,jb} - \be \sum_{\a,\g,\m,\n} V^{(3)}_{ia,\a\g} \G_{\a\m} \G_{\g\n} V^{(3)}_{\m\n,jb} + \OO(\be^2) \  ,
\eeq
where $\b{\G}$ is the rescaled (off-diagonal) correlation defined in Eq.~\eqref{def_Gamma}.
A diagrammatic representation of this is shown in Fig.~(\ref{fig_diags_Jtilde_potts}).

\begin{figure}[t]
\includegraphics[width=8cm]{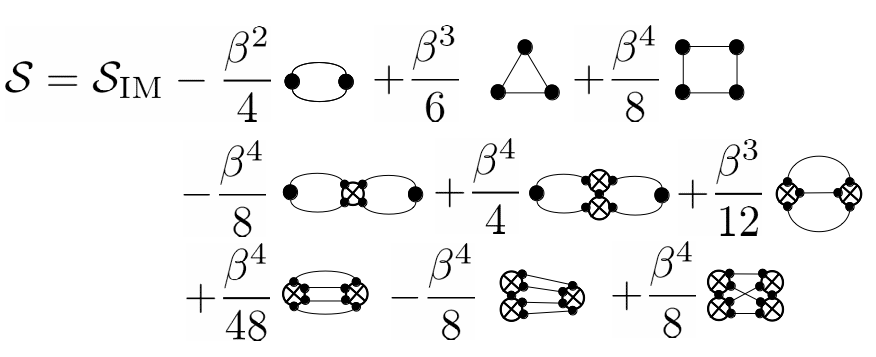}
\caption{Diagrammatic representation of the small correlation expansion of $\SS_{\be}$. Same conventions as in
Fig.~(\ref{fig_diags_Jtilde_potts}).
The first three diagrams are the first terms of the infinite series of ring diagrams, which give $\SS^{\rm ring}$ after resummation.}
\label{fig_S_potts_O4}
\end{figure}

Plugging this result in the definition of $\SS$ we obtain easily its small correlation expansion at order four, which we show only
in a diagrammatic representation in Fig.~(\ref{fig_S_potts_O4}).

\section{Numerical resolution of the matrix $\b{D}$ \label{app_numD}}

Here we present the numerical scheme we have used to compute the matrix  $\b{D}$  in Eq.~\eqref{def_D} of the main text. We rewrite this equation by multiplying it
by $\bL^{1/2}$ to the left and to the right to obtain
\beq
1 = \left(  \mathbb{1}  + \bt{D} + \b{\G} \right)^{-1}_{ii} \quad \forall~ i \ ,
\label{def_D_with-tilde}
\eeq
where $\bt{D}$ and $\b{\G}$ were defined in Sec.~\ref{sec_exp}.
We then define the matrix $\b{X}$,
\beq
\b{X} = \left( \mathbb{1}  + \bt{D} \right)^{-1} \ ,
\eeq
which obviously inherits the diagonal form of $\bt{D}$ i.e. $X_{ij} = X_i \d_{ij}$.
We factorize $\b{X}$ in Eq.~\eqref{def_D_with-tilde} and obtain:
\beq
\left[ \left(  \mathbb{1}  + \b{X} \b{\G} \right)^{-1} \b{X} \right]_{ii} = 1 \quad \forall~ i \ .
\eeq
In a high-temperature expansion, the lowest order reads $\b{\G}=0$ and we would find $\b{X}= \mathbb{1} $.
We isolate this lowest order result by rewriting the above equation as
\beq \begin{split}
& X_{i}  = 1 - F_i[\b{X}] \ ,\\
& F_i[\b{X}] = \left[ \left( \mathbb{1} + \b{X} \b{\G} \right)^{-1}_{ii} -1 \right] X_i\ .
\end{split}\eeq
We solve this equation iteratively by the following procedure
\beq\begin{split}
& X^{(0)}_i = 1 \ ,\\
& X^{(n+1)}_{i} = \a \left( 1 - F_i[\b{X}^{(n)}] \right) + (1-\a) X^{(n)}_{i} \ ,
\end{split}\eeq
where $\a$ is a damping parameter set to values ranging from $0.1$ to $0.01$ for example, that ensures smooth convergence.
This is iterated until a tolerance of $10^{-10}$ on the variations of $\b{X}$ is reached, and the resulting $\b{X}^{\io}$ is injected into
Eq.~\eqref{def_D_with-tilde} to check the convergence. We have forcefully started the iterations from the small-correlations solution,
which we know is physically plausible, and the mixing parameter $\a$
ensures that the procedure always stays close to a physically plausible solution, avoiding spurious instabilities.
We find that the tolerance we have set on $\b{X}$ is such that  the final equation for the $D_i$ is satisfied for each $i$ with an error smaller than
$10^{-8}$. Note that in the low-temperature phase the presence of $\a$ is important, although its precise value is irrelevant, as long as it is small
enough.

If $\bc$ is not invertible, as it might happen in the low-temperature phase or if the sampling is not good enough, then the initial condition $\b{X}^{(0)}= \mathbb{1}$ will not work, as $ \mathbb{1} +\b{\Gamma}$ is not invertible in the first step of the iterative scheme. One should then start from another initial condition, and we have chosen $\b{X}^{(0)}= \b{L}$ in our numerics. With this initial condition, we could always converge a matrix $\bD$, even when $\bc$ is rank deficient.

\section{High-temperature expansion from the $\e$ expansion \label{app_proof}}

We show in this appendix that the approximation to order  $\e^n$ of the Gibbs free energy $\GG_\e[\b{f},\bJ]$ contains the $n+1$-th order of the high-temperature expansion, in the case of Ising variables. The generalization to Potts variables is straightforward.
In order to simplify the counting in couplings, we multiply $\bJ$ by a factor $ \beta$, and study the small $\be$ expansion.

We have already seen in Eq.~\eqref{TAP} that the approximation to order $\e$ is correct to $\OO(\beta^2)$. Then we have to consider the higher-order diagrams (see for example Fig.~\ref{fig_diag2}). The vertices are independent of the temperature, and the temperature dependence is solely contained in the propagator
\beq\begin{split}
\bG(\be) & = \left( \bL^{-1} - \beta\bJ \right)^{-1} - \bL
= \beta\bL \bJ \bL + \OO(\beta^2)\ .
\end{split}\eeq
Using this, we now prove that at each order $n$ of the $\e$ expansion, the corresponding diagrams are at least $\OO(\beta^{n+1})$, which will show that the orders larger than $n$ cannot contribute to the order $n+1$ of the high temperature expansion. Since our expansion is formally exact (assuming it converges), this will prove directly the result by induction.

Let us assume that we have completed the calculation of $\calG_{\e}$ to order $\e^n$, and that the contribution of order $\e^n$ is at least of order $\OO(\beta^{n+1})$, which is true at order $n=1$.  To compute the order $n+1$, we need $\bch$ to order $\e^n$, $\bch^{(n)}$, since it is multiplied by $\e$ in Eq.~\eqref{eq_integralWett}. Schematically,
\begin{equation}
\bch^{(n)}=\bch^{(0)}\sum_{\{s\}}A_s\prod_{m\geq 1} \left(\frac{\d^2\calG^{(m)}}{\d\b{f}\d\b{f}}\right)^{s_m} \bch^{(0)}\ ,
\end{equation}
where the sum is over all possible $\{s\}=\{s_1,s_2,...\}$ such that $\sum_m m s_m=n$, and $A_s$ is some numerical coefficient. One can show that $\frac{\d^2\calG^{(m)}}{\d\b{f}\d\b{f}}=\OO(\beta^{m+1})$. Indeed, since $\calG^{(m)}=\OO(\beta^{m+1})$, to lowest order in $\beta$ we can write it as
\begin{equation}
\calG^{(m)}=\beta^{m+1} F_m(\bJ,{\bf f})\ ,
\end{equation}
where $F_m(\bJ,{\bf f})$ is a function of order $m+1$ in the elements of $\bJ$. Differentiating it twice with respect to $\b{f}$ will not change the power of $\beta$. We thus see that $\prod_{m\geq 1} (\frac{\d^2\calG^{(m)}}{\d\b{f}\d\b{f}})^{s_m}=\OO(\beta^{n+\sum_m s_m})$ which is at least of order $\OO(\beta^{n+1})$. An example of such a term is a generalization of the second diagram of Fig.~\ref{fig_diag2} when the derivatives with respect to the magnetizations (to obtain $\frac{\d^2\calG}{\d\b{f}\d\b{f}}$) acts on the two different vertices:
\begin{equation}
\frac{\d^2\calG^{(n)}}{\d f_i\d f_j} \ni \gamma^{(n+2)}_iG_{ij}^{n+1}\gamma^{(n+2)}_j\,
\end{equation}
which is of order $\OO(\beta^{n+1})$. Then, performing the integration  Eq.~\eqref{eq_integralWett} corresponds to close such terms by adding an additional $\bG$ (see also discussion in Sec.~\ref{sec_potts}), since
\begin{equation}
\partial_\beta \bG=\bch^{(0)}\bJ\bch^{(0)}\ ,
\end{equation}
and thus increasing its order in $\beta$ by at least one, \eg,
\begin{equation}
\calG^{(n+1)} \ni \sum_{ij}\gamma^{(n+2)}_iG_{ij}^{n+2}\gamma^{(n+2)}_j\,
\end{equation}
which is indeed of order $\OO(\beta^{n+2})$.

This completes our proof.

\section{Resummation of 2-spin diagrams in the RMF approximation \label{app_2spin}}

We demonstrate now how one can resum 2-spin diagrams of the entropy for a Potts model (this calculation can also be found in \cite{BLCC16}).
This will illustrate the ACE procedure described in the introduction, and also allow 
us to have a starting point to the inclusion of the 2-spin diagrams in our RMF calculation.

One starts from a system of 2 Potts variable $\b{\s}_i$ and $\b{\s}_j$ with $q$ states, the partition function of which is thus
\beq
Z^{(2)}_{ij}[\b{h},\b{H},\b{J}] = \sum_{a=1}^q \sum_{b=1}^q \exp \left( h_a + H_b + \be J_{ab} \right) \  ,
\eeq
where $\b{h} = \left( h_1 , \ldots, h_q \right)$ are the fields acting on the first variable, $\b{H} = \left( H_1 , \ldots , H_q \right)$ the fields acting on the second one, and $\b{J} = \left( J_{12} , \ldots , J_{(q-1)q} \right)$ are the couplings acting on the pair of variables. We fix the Potts gauge by setting $h_q=H_q=J_{aq}= J_{qb} =0$ for all $a,b=1 \ldots q-1$ which leads to
\beq
Z^{(2)}_{ij}[\b{h},\b{H},\b{J}] = 1 + \sum_{a=1}^{q-1} e^{h_a} + \sum_{b=1}^{q-1} e^{H_b} + \sum_{a=1}^{q-1} \sum_{b=1}^{q-1} e^{h_a} e^{\be J_{ab}} e^{H_b} \  ,
\eeq
so that unless specified otherwise the summations over color indices will now run from $1$ to $q-1$ only. We define reduced variables
\beq
x_a = e^{h_a} \  , ~ y_b = e^{H_b} \  , ~ \chi_{ab} = e^{\be J_{ab}} \  ,
\eeq
and perform the Legendre transform with respect to $\b{h}$, $\b{H}$ and $\b{J}$ simultaneously. We call $\b{f}$, $\b{F}$ and $\b{p}$ the conjugated variables, and the optimization equations are found to be
\beq\begin{split}
& f_a = \frac{x^*_a \left( 1 + \sum_b \chi^*_{ab} y^*_b \right)}{Z^{(2)}_{ij}[\bh^*,\b{H}^*,\bJ^*]} \  , \quad 
 F_b = \frac{y^*_b \left( 1 + \sum_a x^*_a \chi^*_{ab} \right)}{Z^{(2)}_{ij}[\bh^*,\b{H}^*,\bJ^*]} \  , \quad \\
& p_{ab} = \frac{x^*_a \chi^*_{ab} y^*_b}{Z^{(2)}_{ij}[\b{h}^*,\b{H}^*,\bJ^*]}  \  .
\end{split}\eeq
We easily find the value of the partition function by tracing over the color variables and combining the resulting equations
\beq
Z^{(2)}_{ij}[\b{h}^*,\b{H}^*,\bJ^*] = \frac 1{ 1 + \sum_{ab} p_{ab} - \sum_a f_a - \sum_b F_b} \  ,
\eeq
which we abbreviate by $Z^*$ in the following.
Replacing in the optimization equations leads to the solution for the optimal fields and couplings
\beq\begin{split}
& x_a^* = Z^* \left( f_a - \sum_b p_{ab} \right) \  , \quad y_b^* = Z^* \left( F_b - \sum_a p_{ab} \right) \  , \\
& \chi^*_{ab} = \frac{ p_{ab}}{ Z^* \left( f_a - \sum_b \be p_{ab} \right) \left( F_b - \sum_a \be p_{ab} \right)} \  .
\end{split}\eeq
And the entropy is obtained through
\beq
\SS^{(2)}_{ij} = \ln Z^* - \sum_a f_a \ln x^*_a - \sum_b F_b \ln y_b^* - \sum_{a,b} p_{ab} \ln \chi^*_{ab} \  .
\eeq
This depends only, because of our gauge choice, on the colors $1 \cdots q-1$.
The best way to obtain a compact and symmetric result is to define the objects
\beq\begin{split}
& f_q = 1 - \sum_a f_a \  , \quad F_q = 1 - \sum_b F_b \  , \\
&  p_{aq} = f_a - \sum_b \be p_{ab} \  , \quad  p_{qb} = F_b - \sum_a \be p_{ab} \  , \\
&  p_{qq} = 1 +  \sum_{a,b} p_{ab}  - \sum_a f_a - \sum_b F_b \  .
\end{split}
\label{conservation}
\eeq
In a functional sense, they must be understood as functions of the free variables $\b{f},\b{F}$ and $\b{p}$, 
but when considering the data, these relations only express the conservation: the first variable can only
assume one color, leading to $\sum_{a=1}^q f_a = 1$ and similar relations. Replacing these formulas in the
expression of the entropy directly leads to the result (after replacing $p_{ab}$ by $p_{ia,jb}$, $f_a$ by $f_{ia}$, 
and $F_b$ by $f_{jb}$):
\beq
\SS^{(2)}_{ij} = - \sum_{a=1}^q \sum_{b=1}^q p_{ia,jb} \ln \left( p_{ia,jb} \right) \  .
\eeq
As stated in the introduction, this result also incorporates the contributions coming from the two variables 
considered independently, so that the contribution coming solely from the interactions between the two 
variables (often call the excess entropy) is given by
\beq \begin{split}
\D \SS^{(2)}_{ij} = & ~ - \sum_{a=1}^q \sum_{b=1}^q  p_{ia,jb} \ln \left( \frac{p_{ia,jb}}{f_{ia} f_{jb}} \right) \  .
\end{split}\eeq
The total entropy for a system of $N$ Potts variables is given by the independent model entropy
$\SS_{\rm IM}$, plus the summation over all pairs of variables of the excess entropy of the pair, $\D \SS^{(2)}_{ij}$.
We find the result
\beq\begin{split}
& \SS^{\rm 2spin}[\b{f},\b{p}] = \SS_{\rm IM}[\b{f}]  - \sum_{i<j} \sum_{a,b=1}^q p_{ia,jb} \ln \left( \frac{p_{ia,jb}}{f_{ia} f_{jb}} \right) \  .
\end{split}
\label{entropy_2spin}
\eeq
The excess entropy coincides with the well known mutual information, that was used for example 
in the bio-informatics community before the introduction
of direct correlation methods, such as the one we developed in this paper, or such as DCA.
The optimal couplings are found by taking a derivative with respect to $p_{ia,jb}$ with $i<j$ and $a$ and $b$ in \mbox{$\{1 , \ldots ,q-1\}$}, 
taking care that the variables depending on the $q$-th color are in fact functions of the others through Eq.~\eqref{conservation}.
The result is (still for $i<j$ and $a,b \le q-1$):
\beq\begin{split}
J^{\rm 2spin}_{ia,jb} = & ~ \ln \left( \frac{p_{ia,jb} ~ p_{iq,jq}}{p_{ia,jq} ~ p_{iq,jb}} \right) \  .
\end{split}
\label{coupling_2spin}
\eeq
Equations~(\ref{entropy_2spin}--\ref{coupling_2spin}) solve the inference problem, within
the approximation that all pairs of variables interact independently from the others. To compare with the diagrams 
of the small correlation expansion, one has to make the replacement $p_{ia,jb} \to f_{ia} f_{jb} + L_{ia,jb} + \t{c}_{ia,jb}$ and
expand in powers of $\bt{c}$. \\

In order to combine this resummation with our RMF approximation, we need to
identify the two-spin diagrams in the expression of the RMF entropy in Eq.\eqref{RMF_entropy_potts} and substract them before adding 
the entropy of the two-spin model in Eq.~\eqref{entropy_2spin}, in order to avoid double counting. In other words we will have
the expression
\beq
\SS^{\rm RMF+2spin} = \SS^{\rm RMF} + \sum_{i<j}  \D \SS_{ij} - \SS^{\rm dc} \  ,
\label{entropy_RMF+2spin}
\eeq
where $\SS^{\rm dc}$ is the sum of all two-spin diagrams contained in $\SS^{\rm RMF}$. To compute
the diagrams in $\SS^{\rm dc}$, we use the same method than for the two-spin resummation, and consider a system 
of two variables $i$ and $j$ only, with $i \ne j$. 
The matrices involved in the calculations will be of size $2(q-1) \times 2(q-1)$ 
with a structure of four $(q-1) \times (q-1)$ blocks.
We define the blocks of the $\bL$, $\b{D}$ and $\b{\G}$ matrices to be
\beq
\bL = \begin{pmatrix} \bL_i & \b{0} \\ \b{0} & \bL_j  \end{pmatrix} \  , 
~  \b{D} = \begin{pmatrix} \b{D}_i & \b{0} \\ \b{0} & \b{D}_j  \end{pmatrix} \  , ~ 
\b{\G} = \begin{pmatrix} \b{0} & \b{\G}_{ij} \\ {}^t \b{\G}_{ij} & \b{0} \end{pmatrix} \  .
\eeq
and a similar definition for the rescaled $\bt{D}$ matrix defined in Eq.~\eqref{D_rescaled}, with $\bt{D}_i$ and $\bt{D}_j$ blocks
This allows us to solve the equation defining the $\b{D}$ matrix in Eq.~\eqref{eq_DPotts} in terms of $\b{D}_i$ and 
$\b{D}_j$. To do so we first remark that we have in our two-spin case
\beq
 \left( \mathbb{1} + \bt{D} \right)^{-1} \b{\G} = \begin{pmatrix}  
\b{0} & \b{A}  \\
\b{B} & \b{0}
\end{pmatrix} \  ,
\label{ass_matrix_def}
\eeq
where
\beq
\b{A} = \left( \Id + \bt{D}_i \right)^{-1}   \b{\G}_{ij} \ , \quad \b{B} = \left( \Id + \bt{D}_j \right)^{-1}  ~ {}^t \b{\G}_{ij}  \  .
\eeq
The equation defining the $\b{D}$ matrix involves $\left( \Id + \bt{D} +  \b{\G} \right)^{-1}$ which is expanded formally as
\beq \begin{split}
\left( \Id + \bt{D} +   \b{\G} \right)^{-1} & = \left( \Id + \left( \Id + \bt{D} \right)^{-1}   \b{\G} \right)^{-1} \left( \Id + \bt{D} \right)^{-1} \\
& = \sum_{k=0}^{+\io} (-1)^k \begin{pmatrix}  
\b{0} & \b{A}  \\
\b{B} & \b{0}
\end{pmatrix}^k \left( \Id + \bt{D} \right)^{-1} \  .
\end{split}\eeq
We see that the odd terms in this summation will be zero on the diagonal blocks, so that 
they do not contribute to the equation defining $\b{D}$. We are left with the even powers, which give after resummation
\beq 
\begin{array}{ll}
& \left( \Id + \bt{D} +  \b{\G} \right)^{-1}_{ia,ib} =  \left[ \left( \Id -  \b{A} \b{B} \right)^{-1} \left( \Id + \bt{D}_i \right)^{-1} \right]_{ab} \  , \\
& \\
&  \left( \Id + \bt{D} +  \b{\G} \right)^{-1}_{ja,jb} = \left[ \left( \Id -  \b{B} \b{A} \right)^{-1} \left( \Id + \bt{D}_j \right)^{-1} \right]_{ab} \  .
\end{array}
\eeq
The definition of $\b{D}$ in Eq.~\eqref{eq_DPotts} gives thus
\beq \begin{split}
& \left( \Id + \bt{D}_i \right)^{-1}  = \Id - \b{A} \b{B} \  , \\
& \left( \Id + \bt{D}_j \right)^{-1}  = \Id - \b{B} \b{A} \  .
\end{split}
\label{property_Di}
\eeq
These equations are solved after algebraic manipulations by
\beq\begin{split}
& \bt{D}_i = \frac 12 \left[ \left( \mathbb{1} + 4   ~ \b{\G}_{ij} ~ {}^t \b{\G}_{ij} \right)^{1/2} - \mathbb{1} \right] \  , \\
& \bt{D}_j = \frac 12 \left[ \left( \mathbb{1} + 4  ~  {}^t \b{\G}_{ij} ~ \b{\G}_{ij} \right)^{1/2} - \mathbb{1} \right] \  .
\end{split}\eeq
Note that the matrix square root is always well-defined since $\mathbb{1} + 4  \b{\G}_{ij} ~ {}^t \b{\G}_{ij}$ is symmetric positive definite.
We calculate now the different terms in the expression of the entropy. The trace log term can be rewritten as
\beq\begin{split}
& \frac 12 \Tr \left[ \ln \left( \b{c} + \b{D} \right) - \ln \bL \right] = \frac 12 \Tr \ln \left( \mathbb{1} + \bt{D} +  \b{\G} \right) \\
& = \frac 12 \Tr \ln \left( \mathbb{1} + \bt{D} \right) + \frac 12 \Tr \ln \left( \mathbb{1} +  \left( \mathbb{1} + \bt{D} \right)^{-1} \b{\G} \right) \  .
\end{split}
\label{trace-log-term_2spin}
\eeq
Expanding the logarithm in Eq.~\eqref{trace-log-term_2spin}, we see again that the odd powers of the matrix defined in Eq.~\eqref{ass_matrix_def} 
will be zero on the diagonal blocks, so that 
they do not contribute to the trace. We are left with the even powers, which give after resummation and using the cyclicity of the trace
\beq \begin{split}
& \frac 12 \Tr \ln \left( \mathbb{1} + \left( \mathbb{1} + \bt{D} \right)^{-1} \b{\G} \right)  = \frac 12  \Tr \ln \left( \Id - \b{A} \b{B} \right)  \  .
\end{split} \eeq
The other trace log term is easily simplified in
\beq
\frac 12 \Tr \ln \left( \mathbb{1} + \bt{D} \right) = \Tr \ln \left( \Id + \bt{D}_i \right) \  .
\eeq
Eq.~\eqref{property_Di} allows us to simplify the expressions,
so that we have finally the double counting entropy
\beq\begin{split}
\SS^{\rm dc}  = & ~ \frac 12 \sum_{i<j} \Tr \ln \left( \frac{\Id + \left[ \Id + 4   \b{\G}_{ij} {}^t \b{\G}_{ij} \right]^{1/2}}{2} \right) \\
& - \frac 12 \sum_{i<j} \Tr \left( \left[ \Id + 4   \b{\G}_{ij} {}^t \b{\G}_{ij} \right]^{1/2}- \Id \right)
\end{split}\eeq
Since this expression cancels when $\b{\G}$ is set to zero, we see that this entropy does not over-count
diagrams coming from the independent model entropy. 
The optimization of the functional $\SS^{\rm RMF+2spin}$, given in Eq.~\eqref{entropy_RMF+2spin}, over $\b{p}$ will lead to the equation
for the inferred couplings. \\

Note that when the $\b{D}$ matrix is neglected and $\SS^{\rm ring}$ is considered instead of $\SS^{\rm RMF}$, 
the double-counting entropy of ring diagrams is different and we find instead
\beq
\SS^{\rm dc (ring)} = \frac 12\sum_{i<j} \Tr \ln \left( \Id -   \b{\G}_{ij} {}^t \b{\G}_{ij} \right) \  .
\eeq
This procedure of resummation of $k$-spin diagrams can be pursued at least for $k=3$ in the zero-magnetization case in the Ising case
\cite{SM09}.

\end{document}